\documentclass{aa}  

\usepackage{graphicx}
\usepackage{txfonts}
\usepackage[colorlinks]{hyperref}

\def\aap{A\&A}
\def\apj{ApJ}
\def\apjs{ApJS}

\def \hi {\ion{H}{i}}
\def\h2{H$_2$}

\def\kms{km\,s$^{-1}$}

\def\deg{\hbox{$^\circ$}}
\def\arcmin{\hbox{$^\prime$}}
\def\arcsec{\hbox{$^{\prime\prime}$}}

\def\fdg{\hbox{$.\!\!^\circ$}}

\begin{document}

\title{Are observed \hi\ filaments turbulent illusions or density structures? }

   \subtitle{Velocity caustics:  facts and fakes}

   \author{P.\ M.\ W.\ Kalberla \inst{1} \and U.\ Haud \inst{2} }

\institute{Argelander-Institut f\"ur Astronomie,
           Auf dem H\"ugel 71, 53121 Bonn, Germany \\
           \email{pkalberla@astro.uni-bonn.de}
           \and
           Tartu Observatory, University of Tartu,
           61602 T\~oravere, Tartumaa, Estonia }

   \authorrunning{P.\,M.\,W. Kalberla \& U.\ Haud } 

   \titlerunning{\hi\ filaments}

   \date{Received 17 December 2019 / Accepted 21 April 2020}

  \abstract 
{The interstellar medium is affected by turbulence, and observed
  \hi\ structures in channel maps are shaped by turbulent motions. It is
  taken for granted by a few theoreticians that observed \hi\ structures
  do not represent real density enhancement, rather velocity caustics caused by
  velocity crowding. This interpretation has been questioned, and 
  the objections    have led to heated debates.}
{To settle the discussion we verify theoretical key parameters by using the
  Effelsberg Bonn \hi\ Survey (EBHIS) observations.}
{We apply unsharp masking to determine filamentary \hi\ structures at
  high spatial frequencies. In addition we use Gaussian parameters to
  distinguish the cold neutral medium (CNM) from observed \hi\ column
  densities. We compare power spectra and spatial distributions of dust
  and \hi\ column densities in order to distinguish CNM and multiphase column
  densities at various velocity widths. }
{Observations contradict the Velocity Channel Analysis (VCA) postulate
  that the spectral index should steepen with the width of the velocity
  window. We instead find that the thin slice spectral index depends
  strongly on the \hi\ phase composition. Multiphase power spectra
    are steeper for regions with cold gas.  VCA contradicts such \hi\ phase
  dependences on the power distribution. Separating the CNM we find
  that the power spectra are significantly flatter than those for the
  multiphase \hi\ composite. We observe excess CNM power for small-scale
  structures originating from cold dust-bearing filaments that are
  embedded in the CNM. Spectral indices for narrow channel widths depend
  on the Doppler temperature of the \hi\ gas. In the presence of enhanced
  small-scale \hi\ structure the far-infrared emission from dust is also
  enhanced. }
{Small-scale cold filamentary \hi\ structures are predominantly caused by
  density enhancements due to phase transitions rather than by 
  velocity caustics.  }

  \keywords{Turbulence -- ISM: clouds -- ISM:  structure -- (ISM:)  dust,extinction}
  \maketitle
%

\section{Introduction}
\label{Intro}

Density distribution and motion of the interstellar medium (ISM) are
affected by many factors, most notably Galactic dynamics, but  
it has also been suggested that turbulence is   important for all ISM constituents on
all scales \citep{Armstrong1995}. The ISM contains ionized, molecular,
and atomic components; here we are concerned with the neutral hydrogen
(\hi) which fills a large fraction of the Milky Way disk and halo.  
\hi\ is easily observable and is therefore an ideal tracer for various
ISM structures. All-sky surveys, like the combined Effelsberg and
Parkes \hi\ survey \citep[HI4PI,][]{Winkel2016c} or the Galactic Arecibo
L-Band Feed Array Survey \citep[GALFA-\hi,][]{Peek2018}, disclose a
wealth of structures in data cubes that are organized in channel maps
with position-position-velocity (PPV) coordinates. Comparing the new
high resolution surveys with the older low resolution
Leiden-Argentine-Bonn (LAB) data \citep{Kalberla2005}, we find that
the new data contain a striking wealth of previously unknown
filamentary \hi\ structures that are aligned with the magnetic field
and correlated with far-infrared (FIR) emission observed by {\it
  Planck}, see \citet{Clark2014,Clark2015}, \citet{Kalberla2016},
\citet{Peek2018}, \citet{Clark2019}, and \citet{Clark2019b}.
The access to the \hi\ survey data is easy, but the interpretation is
hampered by the fact that these data do not contain the full 3D phase
space information. They offer only projections with information for
two perpendicular spatial coordinates  and a single one along the line
of sight in velocity.

Restrictions for the turbulence analysis of such PPV data have been
considered by \citet{Lazarian2000} for an isothermal medium under the
assumption that the velocity field is uncorrelated with the density
field. They have argued that 3D power laws for the density field derived
from PPV column density maps should depend on the width of the velocity
slice used. Intensity fluctuations in thin velocity slices are affected
by both density and velocity fluctuations.  In particular, when the
density spectrum is long-wave dominated, the \hi\ intensity spectrum
does not reflect the density statistics. However, in the limit of very
thick slices the velocity effects on the projected intensity
distribution integrate out.  \citet{Lazarian2000} suggest that the
velocity fluctuations for thin slices make the spectra of emissivity
more shallow, which creates many structures in
position-position-velocity (PPV) space that might be identified as
clouds (see their Sect. 7).  Such spurious structures are be caused by
velocity crowding along the line of sight and are called velocity
caustics. \citet{Lazarian2018} claim that filamentary structures in thin
channel maps, mentioned in the previous paragraph and discussed in
detail by \citet{Clark2014,Clark2015}, are predominantly velocity
caustics, thus not real but an illusion caused by turbulence.

\citet{Clark2019} object against this interpretation. They consider the
spatial correlation between small-scale \hi\ filaments and {\it Planck}
857 GHz emission \citep{Planck2018IV} as an indication that filamentary
\hi\ structures are true density structures in the ISM. From the fact
that the best FIR--\hi\ correlation as well as the highest FIR-to-\hi\ ratio
is found for cold filamentary \hi\ structures, they infer that the
enhanced FIR emission is associated with colder, denser phases of the
ISM. Thus, structures must be caused by dust-bearing density structures
and not by velocity caustics. The anisotropic cold \hi\ structures are
most obvious in thin velocity slices and are observable at high spatial
frequencies. Such \hi\ features are washed out in thick velocity slices
where structures are dominated by the more extended and isotropic
WNM. Changes in spectral slope are attributed to changes in the
multiphase composition of the \hi\ rather than to velocity
fluctuations. \citet{Clark2019} conclude that thin slice spectral
indices must be affected by ``small-scale structure and narrow line
widths typical of CNM'' and call for ``a significant reassessment of
many observational and theoretical studies of turbulence in \hi.''

\citet{Yuen2019} comment on \citet{Clark2019}, and argue that thin
channels always have a contribution from velocity fluctuations. They
note that the ``most valuable insight from \citet{Lazarian2000} is the
prediction of the spectral slope change between the thin and thick PPV
slices that is related to the spectral indices of turbulent velocity and
density.'' This is in fact the heart of the velocity channel analysis
(VCA). The theory was extensively tested with numerical data; we refer
to references given by \citet{Yuen2019}. VCA was formulated for the
isothermal gas, but \citet{Yuen2019} argue further that temperature
inhomogeneities would increase the weight of velocity related
contributions to the velocity channel map, keeping the VCA phenomenon of
spectral slope transitions intact. They claim that two-phase medium and
one-phase medium both show the same result. They question further that
the \hi\ filaments are indicative of cold gas and note that the data
presented by \citet{Clark2019} are close to the North Galactic Pole
where ``neither formation of cold \hi\ nor an increase of dust
emissivity is likely.'' Last not least, \citet{Yuen2019} give counter
examples for a particular region and question whether the relation
between FIR emission and \hi\ column density, derived by
\citet{Clark2019}, is universal.

Summing up, \citet{Clark2019} argue that filamentary
\hi\ structures that are preferentially aligned with the magnetic field
and correlated with FIR emission are cold and dense, initiated by phase
transitions, and thus modifying the thin velocity channel power
index. \citet{Yuen2019} emphasize that temperature inhomogeneities would
not affect VCA; they question whether filaments are cold, and refer to
selection effects that invalidate a general correlation between FIR
emission at 857 GHz and \hi\ column density as used by
\citet{Clark2019}.

Putting these arguments together, we can work out three key questions:
(i) Is VCA applicable to a multiphase medium with temperature
inhomogeneities and embedded cold small-scale structures? (ii) How
cold are \hi\ filaments? and (iii) How close is the correlation between
FIR and \hi\ filaments?

\section{Observations and data reduction }
\label{Observations}

\citet{Clark2019} used Arecibo data. For a completely independent
analysis we utilized here the Effelsberg Bonn \hi\ Survey (EBHIS). We
used for consistency the same region as \citet{Clark2019} and
\citet{Yuen2019} and focus our analysis on a region with a diameter of
15\deg\ around longitude GAL$ = 35\fdg5$ and latitude GAB$ = 54\deg$ (
RA = $15^{\mathrm{d}} 30^{\mathrm{m}} 07^{\mathrm{d}}$, DEC = 23\deg
06\arcmin 00\arcsec). For apodization we apply a cosine taper \citep[a
  Tukey window,][]{Harris1978} and taper with a half-period of 15\degr,
weighting smoothly from one to zero for a radius $ 15\degr < R < 30\degr
$. We calculate power spectra for velocities $|v_{\mathrm{LSR}}| \leq
25$ \kms\ with channel widths of $\Delta v_{\mathrm{LSR}} \leq 51$
\kms\ in the multipole range $l < 1023$ \citep{Kalberla2019}. To
determine the small-scale spatial structure of the \hi\ distribution at
scales of $R < 0\fdg5 $ we apply unsharp masking (USM)
\citep{Kalberla2016}. The \hi\ data are first smoothed with a Gaussian
beam of 0\fdg5 at full width at half maximum; these smooth data are then
subtracted from the original database to generate the USM data. In
addition, we use a Gaussian decomposition of the HI4PI data on an nside
= 1024 HEALPix grid to get parameters for different phases, and
distinguish between the cold, lukewarm, and warm neutral medium (CNM,
LNM, and WNM, respectively, see \citet{Kalberla2018}).

Our methods and data processing, including deep discussions of
instrumental issues, are detailed in \citet{Kalberla2016},
\citet{Kalberla2018}, and \citet{Kalberla2019}.  We note that the
signal-to-noise ratio for the EBHIS is high enough to warrant a power
analysis without a noise bias. For our extended analysis we use in
general only data with a signal-to-noise ratio higher than three,
corresponding to brightness temperatures $T_\mathrm{B} > 0.3$ K.
Another major advantage of our current analysis is that we use EBHIS
data alone; we do not need to consider complications in data processing
related to different telescopes, as in the case of HI4PI
\citep{Kalberla2019}.

\section{Power spectra }
\label{Power}

The Milky Way \hi\ distribution differs from the simplified case of an
isothermal, homogeneous, and isotropic distribution with turbulent
properties that are random and uncorrelated in density and velocity, as
assumed by \citet{Lazarian2000}. In the first instance the column
density of the observed \hi\ depends strongly on position and radial
velocity;  the contributions from CNM, LNM, and WNM are then expected to
vary greatly. In particular the cold part of the \hi\ distribution,
  represented by the USM and CNM, tends to form filamentary structures
  with strong anisotropies, see Fig. \ref{Fig_Gauss_0}. If the
conjecture by \citet{Clark2019} against caustics is correct, we should
expect an imprint of these conditions to the derived power spectra.


\begin{figure}[th] 
    \centering
    \includegraphics[width=9cm]{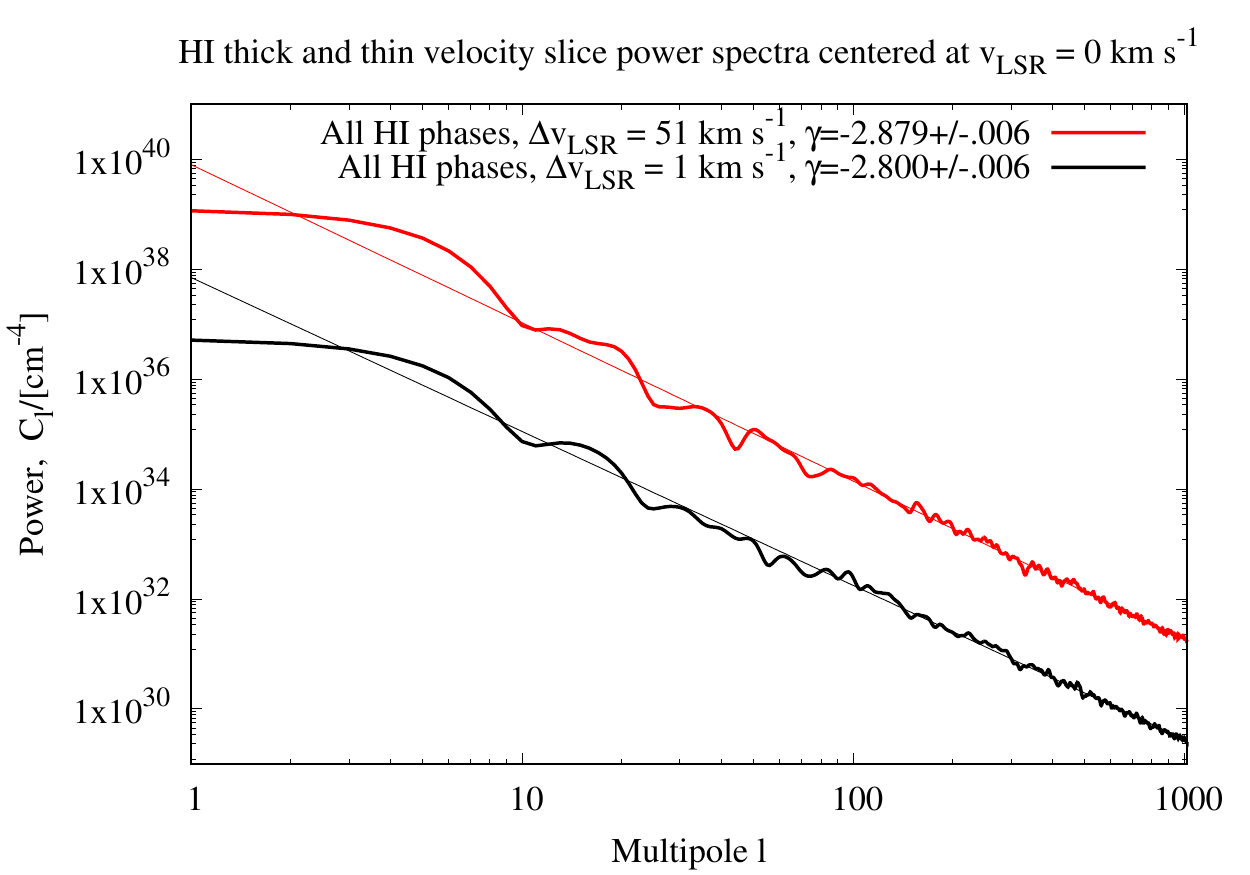}
    \caption{Power spectra of the observed \hi\ column density
      distribution at $v_{\mathrm{LSR}} = 0 $ \kms\ for an instrumental
      channel width of $\Delta v_{\mathrm{LSR}} = 1.29 $ \kms,
      corresponding to the instrumental resolution (black) and for the
      \hi\ integrated in the range $ -25 < v_{\mathrm{LSR}} < 25 $
      \kms\ (red). Power indices from least-squares fits for $l > 10$ are
      indicated; the indicated errors are formal 1$\sigma$ uncertainties.  }
   \label{Fig_Plot_Power}
\end{figure}

\begin{figure}[th] 
    \centering
    \includegraphics[width=9cm]{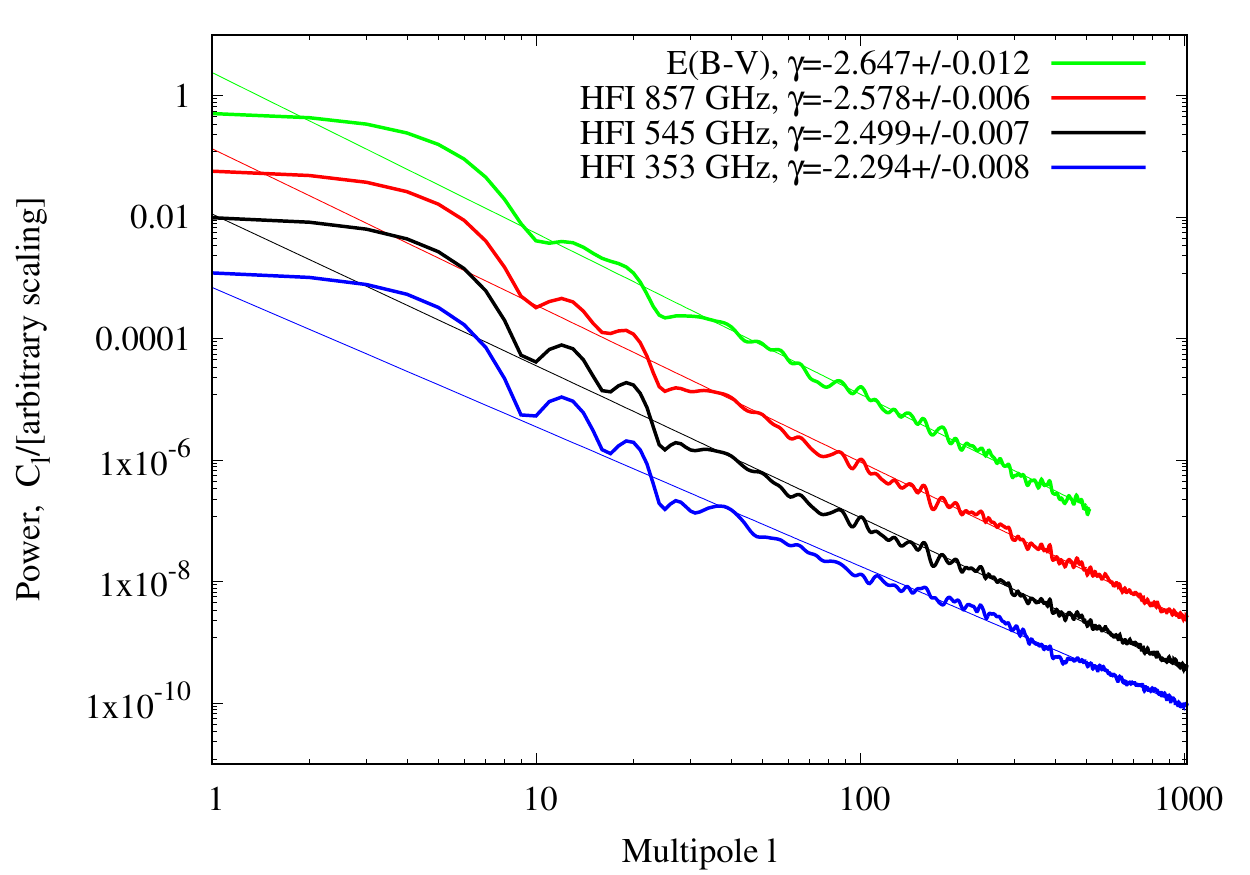}
    \caption{Power spectra for E(B-V) and HFI emission at 857, 545, and
      353 GHz with the corresponding least-squares fits. }
   \label{Fig_HFI}
\end{figure}

Figure \ref{Fig_Plot_Power} shows power spectra for EBHIS data within a
single channel and for column densities integrated over $ -25 <
v_{\mathrm{LSR}} < 25 $ \kms. In both cases the spectra are straight and
well defined for $l \ga 10$; the fits are also well defined with $\gamma
= -2.800 \pm 0.006$ for the single channel and $\gamma = -2.879 \pm
0.006$ for the thick velocity slice. The low power for $l \la 10$
reflects the limitations of our field of view with a radius of $R =
15\deg$, corresponding to a multipole $l \sim 180\deg / 2R = 6 $.  In
all cases we fit power spectra for $l \ge 10$ only.

According to VCA the spectral indices in the limits of thin and thick
velocity slices determine uniquely the spectral indices of the turbulent
velocity and density fields. In the case of sufficiently thick velocity
slices the velocity fluctuations average out, and we measure only
density fluctuations. In addition to density fluctuations, thin velocity
slices also contain velocity fluctuations. These average out on large
scales, hence velocity fluctuations dominate small scales and
accordingly the power spectra are flattened for thin velocity slices.
The difference between thin and thick slice spectral indices can then be
used to determine velocity fluctuations \citep[][Table 1]{Yuen2019}.
Taking the data presented in Fig. \ref{Fig_Plot_Power} at face value, we
conclude that the observed spectral index $\gamma$ steepens with the
width of the velocity slice, as expected from theory; however, the
difference $\delta \gamma_{\mathrm{VCA}} = 0.08$ is far below the
expectations, indicating $ m \sim 0.16 $ for the slope of velocity
structure function. This is only $1/4$ of $m = 2/3$ that would be
expected for Kolmogorov turbulence \citep[][Sect. 5.1]{Lazarian2000}.

\citet{Clark2019} compare \hi\ data with {\it Planck} 857 GHz emission,
assuming that gas and dust are well mixed. If this assumption is valid we
 expect that \hi, FIR, and reddening power spectra are compatible.
We calculate power spectra for {\it Planck} data at 857, 545, and 353
GHz\footnote[1]{HFI\_SkyMap\_857\_2048\_R3.01\_full.fits,
  HFI\_SkyMap\_545\_2048\_R3.01\_full.fits, and
  HFI\_SkyMap\_353-psb\_2048\_R3.01\_full.fits from the {\it Planck}
  Legacy Archive, \url{https://pla.esac.esa.int/}}. In addition we
select E(B-V) reddening data from \citet{Schlegel1998}\footnote[2]{The
  LAMBDA Reddening (E(B-V)) Map,
  \url{https://lambda.gsfc.nasa.gov/product/foreground/fg_sfd_get.cfm}}.
In all cases we use the same field of view and the apodization described
earlier for the \hi. The power spectra, shown in Fig. \ref{Fig_HFI} are
calculated for $l > 10$; the E(B-V) data are only available for $l <
512$. The fits for the spectral indices show a systematical flattening
with decreasing frequency, explainable by changes in the power
distribution. The Galactic residual contribution to {\it Planck} power
spectra is dominant at frequencies $\nu > 217$ GHz, the cosmic microwave
background (CMB) contribution is increasing for lower frequencies, while
the contribution from the cosmic infrared background (CIB) to
Fig. \ref{Fig_HFI} is most significant at 353 GHz
\citep[][Fig. C.1]{Planck2014}.  Comparing power spectra for \hi\ with
spectra for the dust distribution we note that the power spectra for the
dust are well defined, but  in general have significantly flatter power
spectra than the observed \hi\ column densities
(Fig. \ref{Fig_Plot_Power}).


\begin{figure}[th] 
    \centering
    \includegraphics[width=9cm]{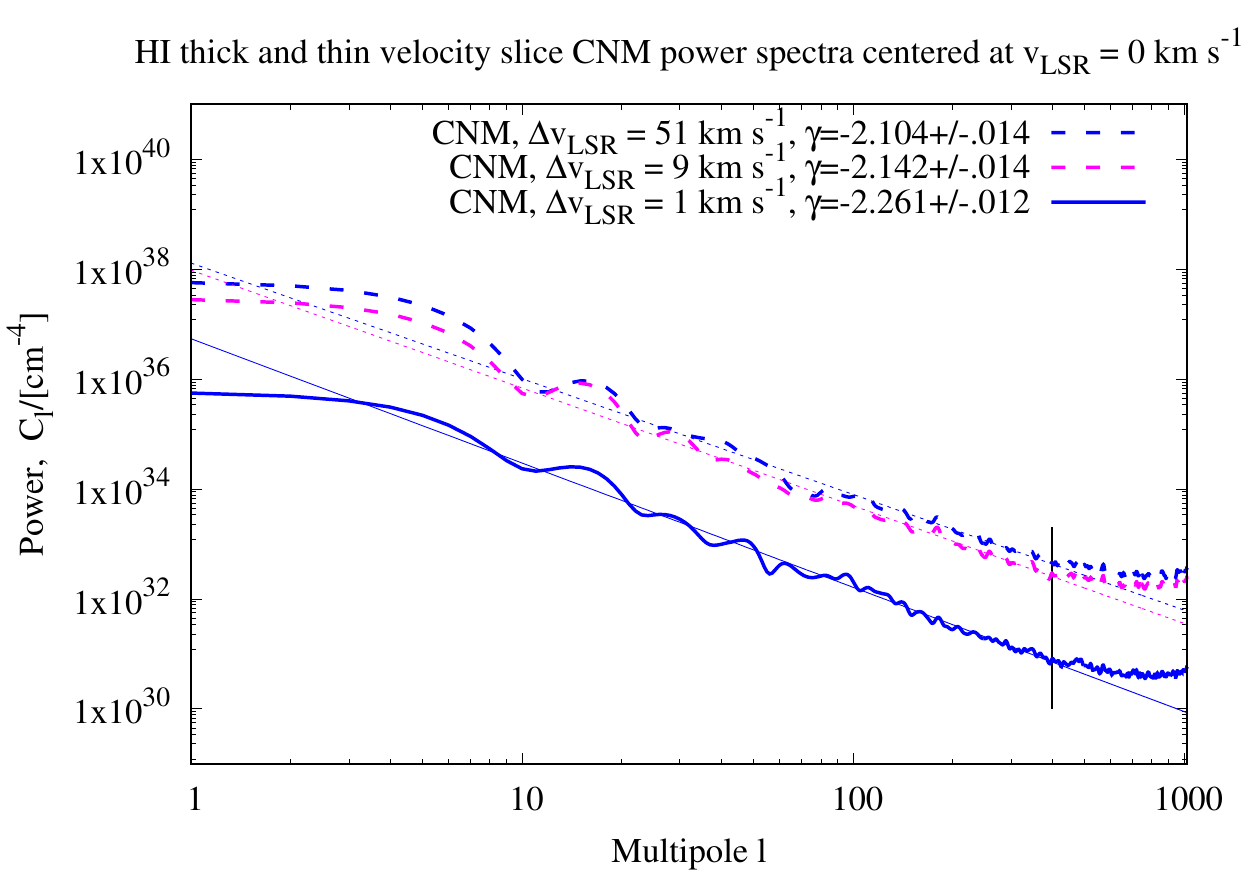}
    \caption{As in  Fig. \ref{Fig_Plot_Power}, but for power spectra of the CNM
      column density distribution at $v_{\mathrm{LSR}} = 0 $ \kms\ for
      $\Delta v_{\mathrm{LSR}} = 1.29 $ \kms\ (blue solid), $ -25 <
      v_{\mathrm{LSR}} < 25 $ \kms\ (blue dashed), and $ -4 <
      v_{\mathrm{LSR}} < 4 $ \kms\ (magenta dashed). Power spectra were
      fit for $10 < l < 400$.}
   \label{Fig_Plot_Power_CNM}
\end{figure}

\citet{Clark2019} argue that dust is correlated best with cold
\hi\ filaments. If their arguments are valid, we   expect  power
spectra for the cold \hi\ to be in better agreement with
Fig. \ref{Fig_HFI} than the spectra from Fig. \ref{Fig_Plot_Power}. To
derive the power distribution of the cold \hi\ we use Gaussian
components for the CNM from \citet{Kalberla2018} and
\citet{Kalberla2019} and repeat the calculation of the thin and thick
slice power spectra for the CNM. The results are given in
Fig. \ref{Fig_Plot_Power_CNM}. Comparing the multiphase power
  spectra from Fig.  \ref{Fig_Plot_Power} with
  Fig. \ref{Fig_Plot_Power_CNM} we find that the CNM power spectra are
   shallower. The power is significantly reduced, but at multipoles $l
  \ga 400$ there is a strong excess. The implication is that the
\hi\ distribution on large scales (low multipoles) is dominated by warm
gas, while the CNM controls small scales, as claimed by
\citet{Clark2019}. The small-scale power excess is strongest for the
thin slice power spectrum. We also note  that we observe at
intermediate multipoles $10 < l \la 400$ a steeper CNM power spectrum in the
case of the thin slice, opposite to VCA expectations
\citep[][Sect. 4.3]{Lazarian2000}. Similar results were reported by
\citet{Kalberla2019}. These authors demonstrate that the power
  excess at high multipoles for HI4PI observations is not
  affected by noise or instrumental uncertainties. To test whether the
  same conditions apply to our current data analysis we calculate for
  comparison a CNM power spectrum for an intermediate channel width of $
  -4 < v_{\mathrm{LSR}} < 4 $ \kms.  Comparing this case with the thick
  velocity slice for $ -25 < v_{\mathrm{LSR}} < 25 $ \kms, we find only
  slight changes in the power spectra caused by the source
  distribution. The CNM has most of its power at velocities close to
  zero. There is a slight steepening, but no significant change in the
  power excess for $l \ga 400$. If this excess  had been caused by
  uncertainties in the data analysis, we should notice an increase of a
  factor of 5.7 (or log(5.7) = 0.75), corresponding to the decrease in
  channel width. This is not observed.  

\begin{figure}[th] 
    \centering
    \includegraphics[width=9cm]{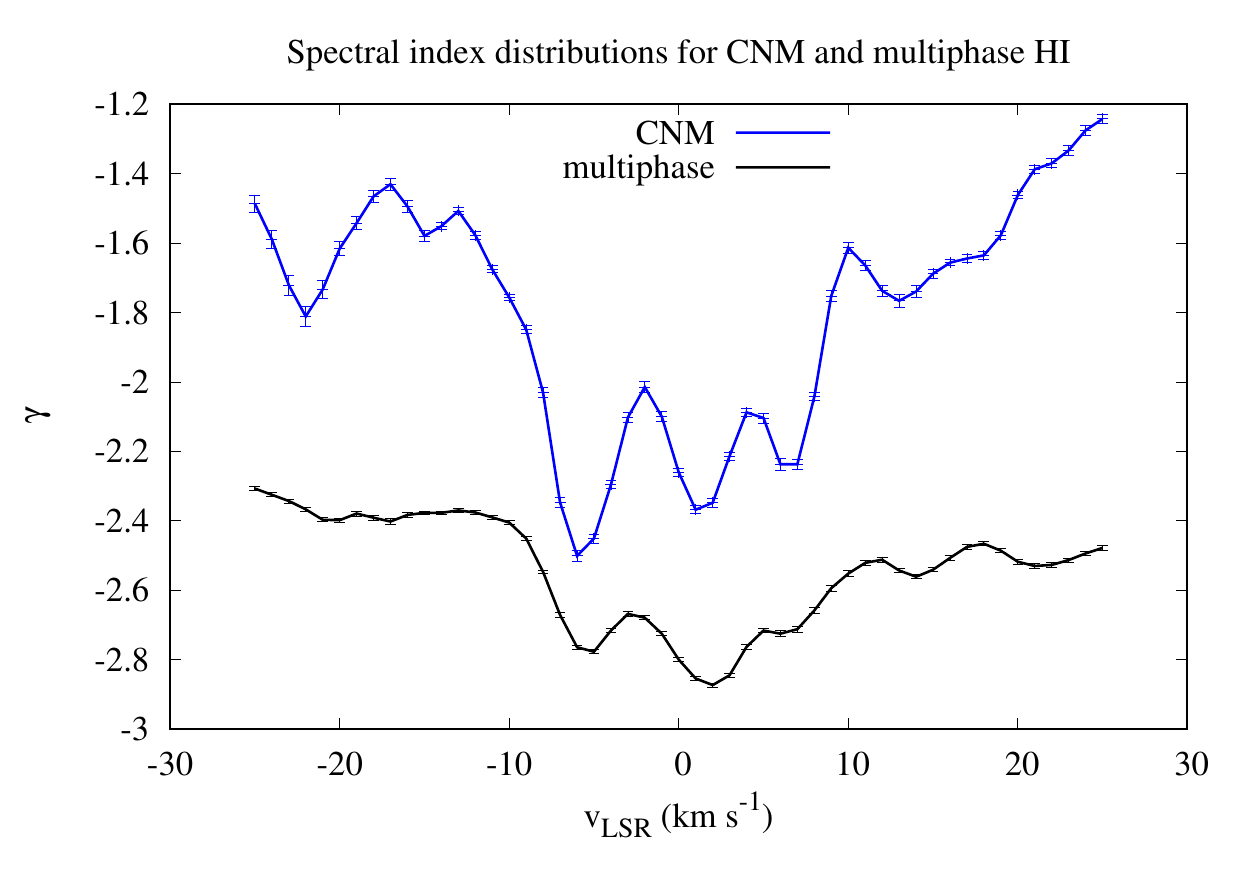}
    \caption{Velocity dependent single-channel power indices $\gamma$
      for the multiphase \hi\ (black) and for the CNM (blue) with formal 1$\sigma$ error bars for the fit.  }
   \label{Fig_Multi_Power}
\end{figure}

The filamentary USM structures used by \citet{Clark2019} are strongly
velocity dependent, we therefore expect that  power spectra for cold
gas should also show some velocity dependences.  To check, whether $\gamma$ is
velocity dependent, we calculate  single-channel power spectra for $ -25
< v_{\mathrm{LSR}} < 25 $ \kms\ for observed multiphase \hi\ column
densities and also for the CNM. The spectral indices are displayed in
Fig. \ref{Fig_Multi_Power}. We find fluctuations up to $\delta \gamma
\la 0.5$ in the case of the multiphase \hi\ and $\delta \gamma \la 1$ in
the case of the CNM. The velocity dependences are large compared to $\delta
\gamma_{\mathrm{VCA}} = 0.08$ from Fig. \ref{Fig_Plot_Power}. 

Figure \ref{Fig_Multi_Power} shows three local minima in $\gamma$. The
multiphase spectral index is steepest at $ v_{\mathrm{LSR}} = 1$ \kms\
and the CNM power index has a local minimum at $ v_{\mathrm{LSR}} = 0 $
\kms.  The second multiphase local minimum occurs at $
  v_{\mathrm{LSR}} = -6 $ \kms. The thin slice power spectrum for the
  CNM is steepest at this velocity, but still shallower than the
  multiphase. Significant fluctuations of the thin velocity slice
  spectral index with velocity are not expected for the multiphase
  \hi\ in the VCA framework. Moreover, the changes in spectral index are
  driven by the CNM and we shall see in Sect. \ref{Visual} that these
  fluctuations are associated with changing anisotropies
  (Fig. \ref{Fig_Gauss_0}). The question arises whether VCA is
  applicable under such conditions. 


\begin{figure}[th] 
    \centering
    \includegraphics[width=9cm]{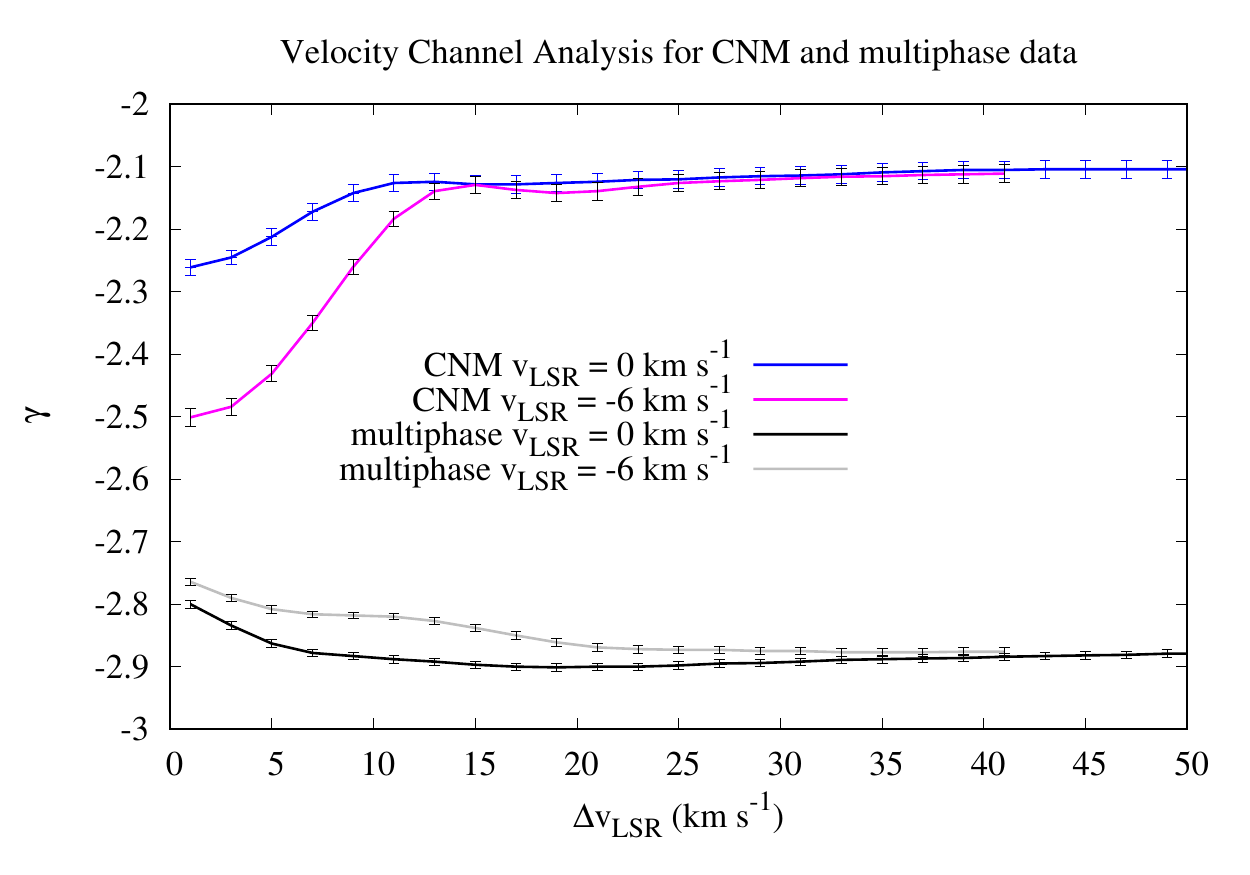}
    \caption{Velocity dependent single-channel power indices $\gamma$
      for the multiphase \hi\ (black) and for the CNM (blue) with formal 1$\sigma$ error bars for the fit.  }
   \label{Fig_Gamma_VCA}
\end{figure}

Next we check the VCA key prediction that the spectral index should
gradually steepen with the transition from thin to thick velocity
slices. A slice is considered to be thick for a width of $\Delta
v_{\mathrm{LSR}} \ga 17$ \kms, as expected for warm \hi\ at a Mach
number of 1 \citep[][Sect. 4.3]{Lazarian2000}. We calculate power
spectra in slices with $1.26 < \Delta v_{\mathrm{LSR}} \la 51$ \kms, a
range that should cover  the transition from thin to thick velocity
slices well. In theory, the VCA predicted spectral steepening with velocity
width should be independent of the center velocity. Observed velocity
dependences  of the thin slices from Fig. \ref{Fig_Multi_Power} may
imply that such dependences can also affect  the VCA and we intend to
check possible dependences of this kind. We use two of the prominent
local minima from Fig. \ref{Fig_Multi_Power} with $ v_{\mathrm{LSR}} =
0$ \kms, and $ v_{\mathrm{LSR}} = -6$ \kms. Figure \ref{Fig_Gamma_VCA}
displays the VCA results for the CNM alone  and for the observed
multiphase \hi.


For the multiphase \hi\ with a center velocity $ v_{\mathrm{LSR}} = 0$
\kms\ we see in Fig. \ref{Fig_Gamma_VCA} that the steepest spectral
index is reached at $\Delta v_{\mathrm{LSR}} \sim 17$ \kms\ but for the
very wide range $\Delta v_{\mathrm{LSR}} \ga 17$ \kms\ the power
spectrum appears to flatten again. In the case  of a center velocity $
v_{\mathrm{LSR}} = -6$ \kms\ the steepening is somewhat stronger, but
still not as significant as predicted by VCA. The result for the
  CNM clearly indicates that VCA is not applicable in this case. The
spectral index flattens progressively with increasing $\Delta
v_{\mathrm{LSR}} $. The velocity dependences, shown in
Fig. \ref{Fig_Multi_Power}, dominate in any case the VCA statistics for
the cold \hi.  We show in Fig. \ref{Fig_Plot_Power_CNM} that
  most of the CNM power is concentrated at low velocities. The
  implication is that the surface filling factor for the CNM is largest
  at these velocities with a dominance of small-scale structures, as
  visible in Fig. \ref{Fig_Multi_Power}. 

We conclude that the objections by \citet{Clark2019} against VCA are
justified. Spectral steepening cannot be explained as a decreasing
contribution of the velocity field, but rather as a shallower power
spectrum for the cold gas in a narrow velocity range. Thus, changes in
spectral index must be caused by temperature fluctuations, indicative
for phase transitions.

\section{Doppler temperatures}
\label{Cold}

In this section we  discuss the second key question of whether
\hi\ filaments are cold, as claimed by \citet{Clark2019}. Filamentary
structures discussed by \citet{Clark2014,Clark2015} are derived by
unsharp masking. Low spatial frequencies are suppressed by this method,
but the calculations are done independently for each channel, hence
velocity dependences of the data (or correlations in velocity) remain
unaltered. A priori, a USM treatment should disclose small-scale
structure in the observed \hi\ distribution regardless of whether the
observed gas is cold or warm. In other words, USM data should show
fluctuations in the CNM and in the WNM at high spatial
frequencies. 


\begin{figure}[th] 
    \centering
    \includegraphics[width=9cm]{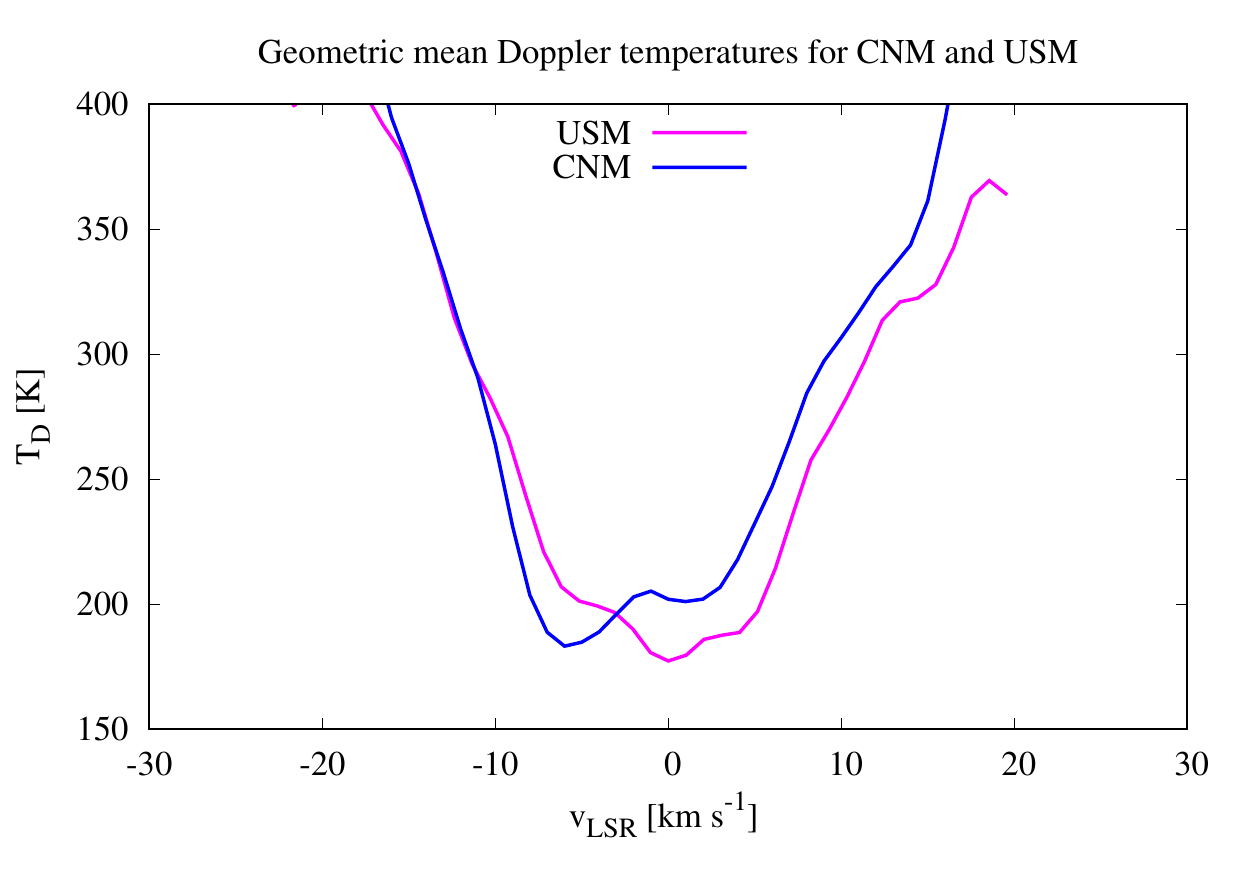}
    \caption{EBHIS harmonic mean Doppler temperatures, determined from
      small-scale USM structures and for the CNM from a Gaussian
      decomposition. USM data for $v_{\mathrm{LSR}} > 20$ \kms\ are
      excluded because of their low significance. }
   \label{Fig_Doppler}
\end{figure}

\begin{figure*}[th] 
   \centering
   \includegraphics[width=5.4cm,angle=-90]{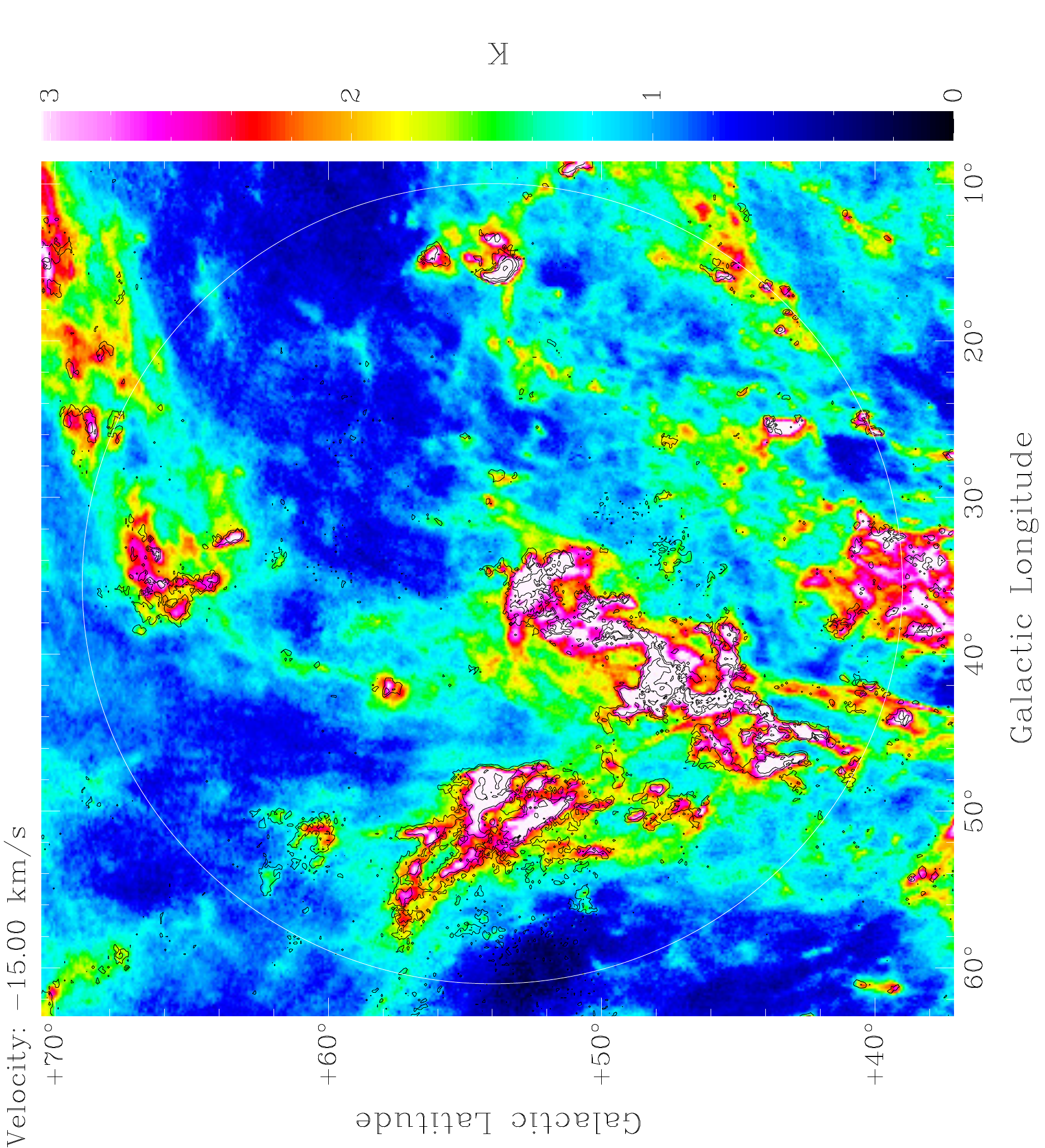}
   \includegraphics[width=5.4cm,angle=-90]{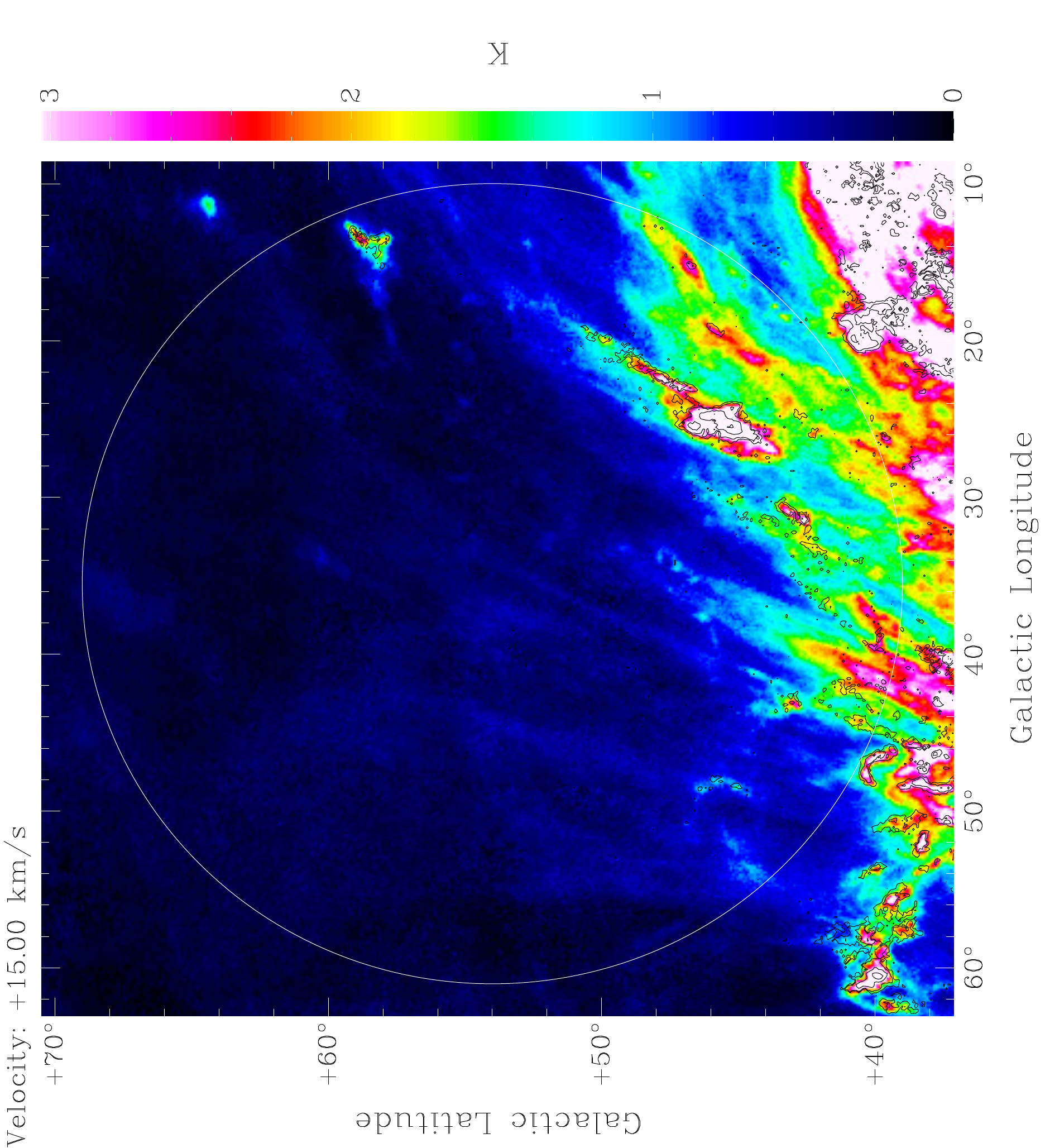}
   \includegraphics[width=5.4cm,angle=-90]{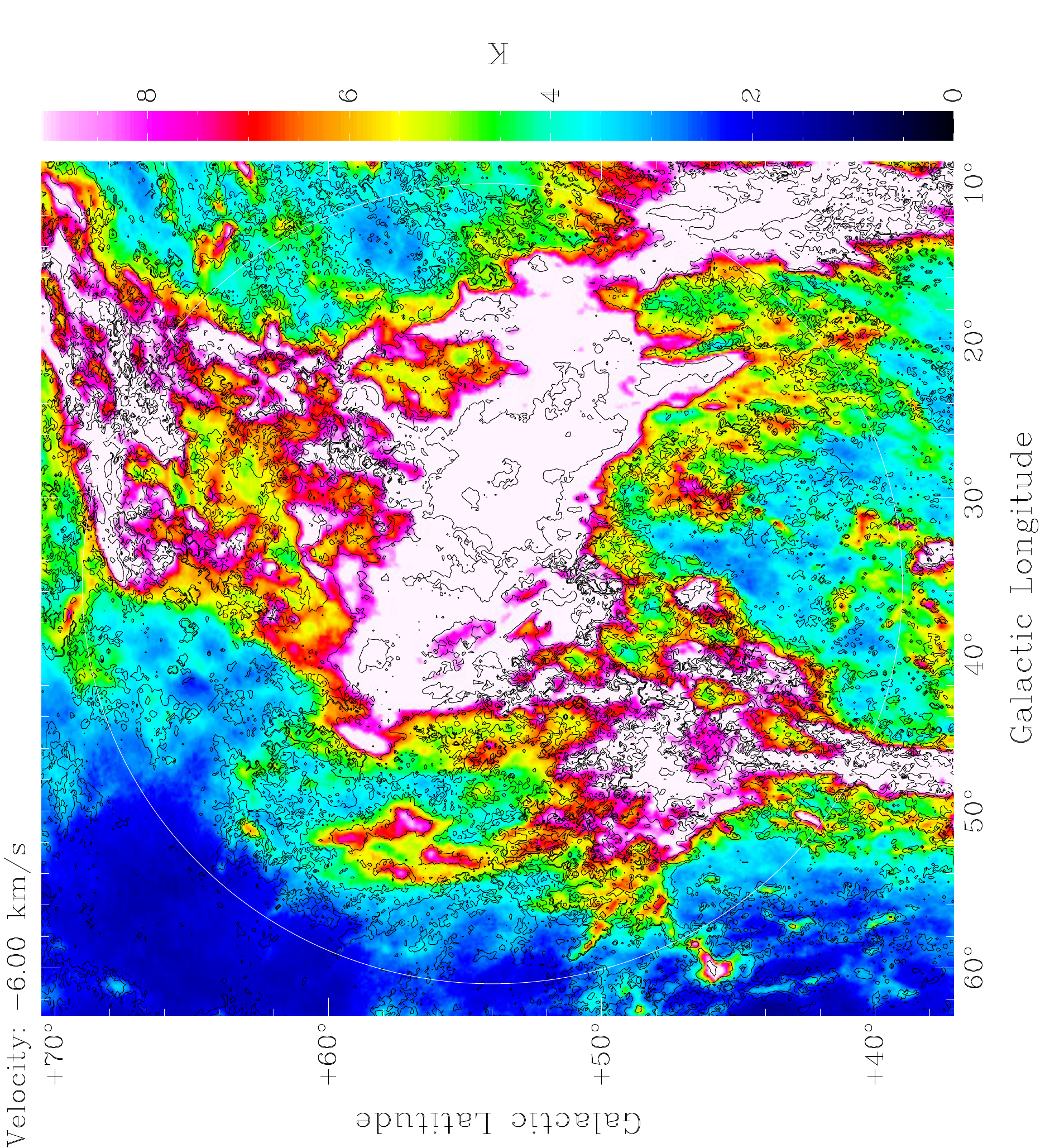}
   \includegraphics[width=5.4cm,angle=-90]{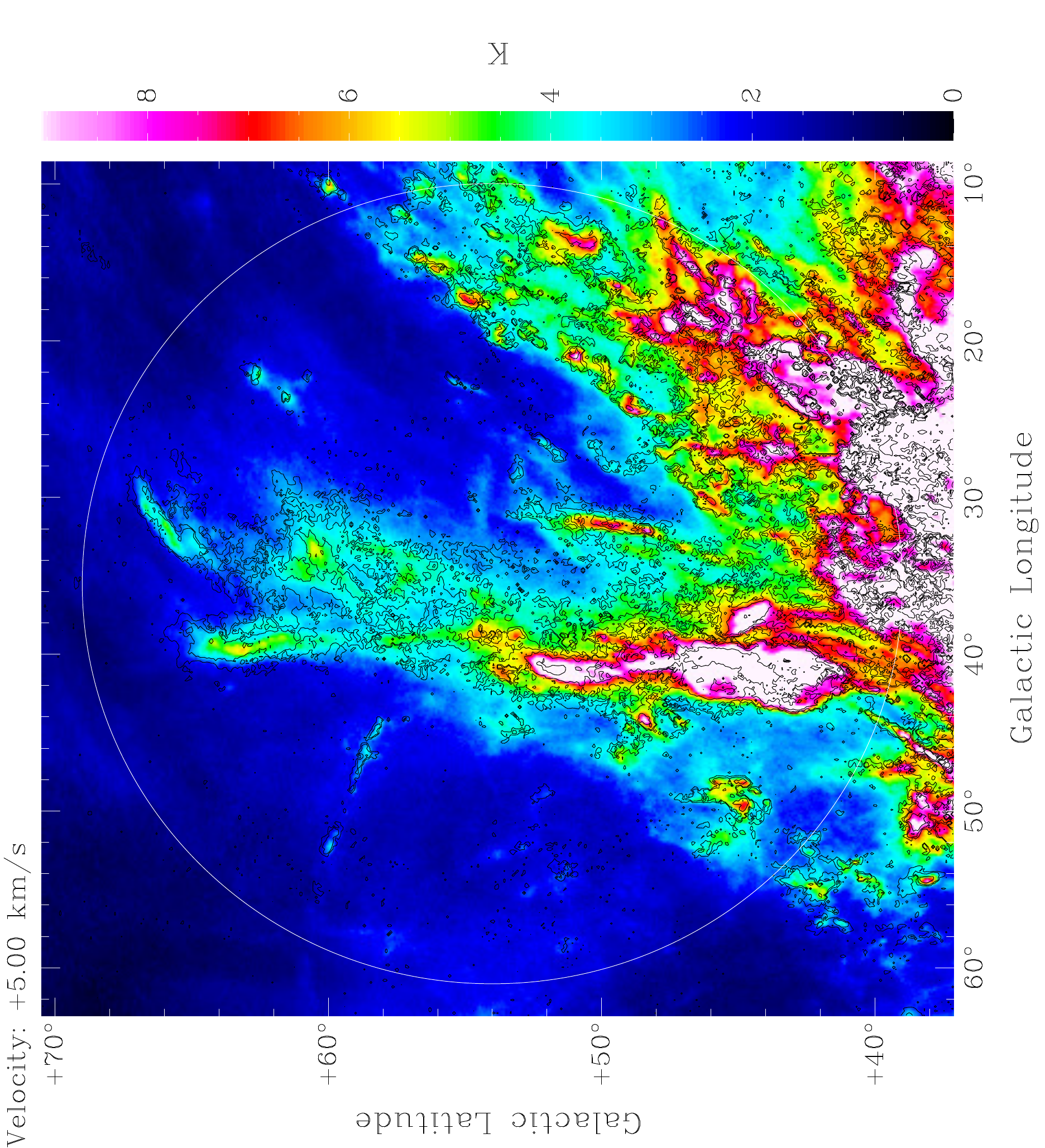}
   \includegraphics[width=5.4cm,angle=-90]{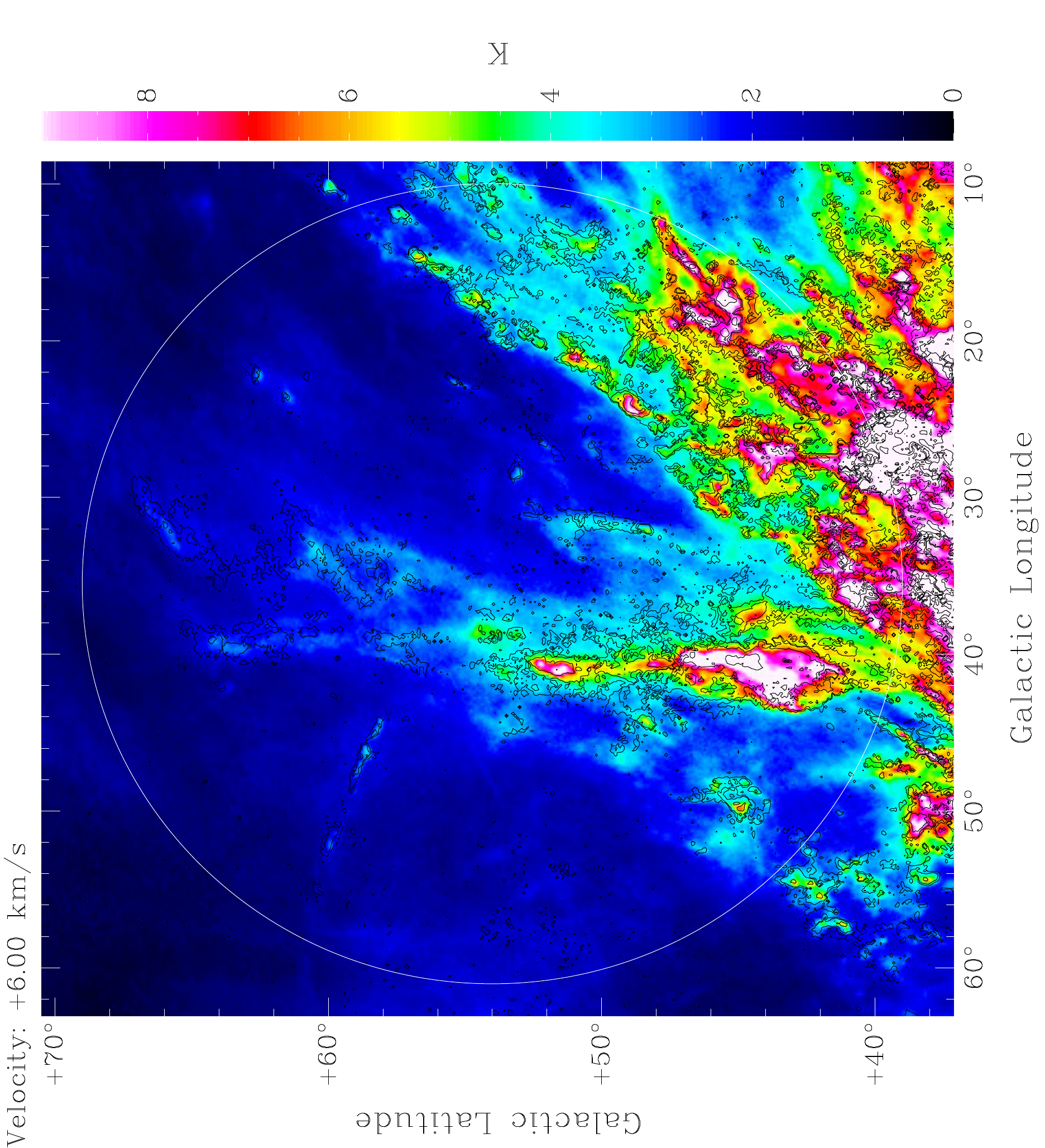}
   \includegraphics[width=5.4cm,angle=-90]{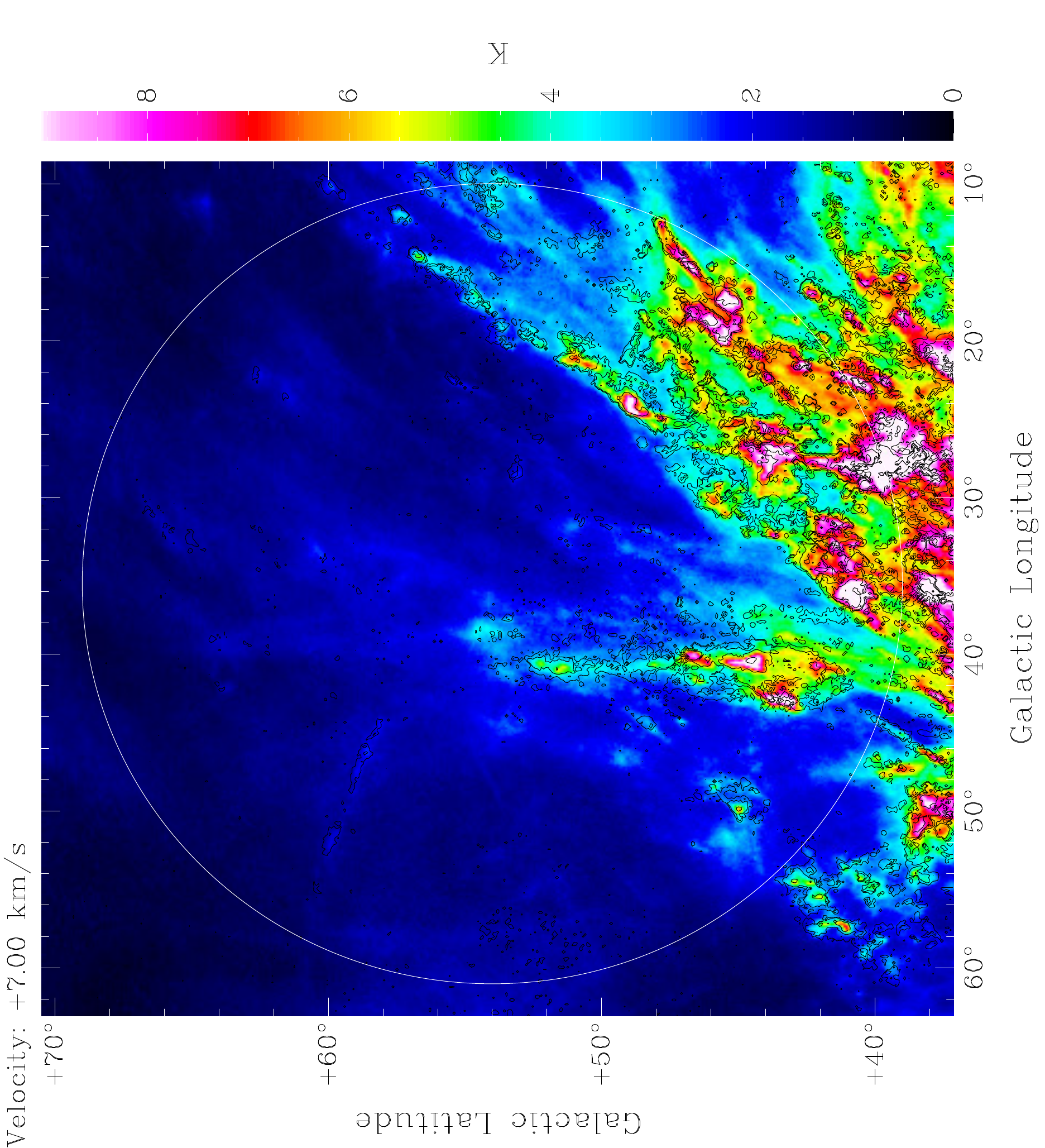}
   \caption{The distribution of CNM brightness temperatures (isophotes
     at $T_{\mathrm {B,CNM}} = 0.3$, 1, 3, and 9 K) compared to multiphase
     brightness temperatures $T_\mathrm{B}$ (color coded) for the
     central part of the observed field. We display the distributions at
     $v_{\mathrm{LSR}} = -15$, +15, and -6 \kms\ on top and for
     $v_{\mathrm{LSR}} = +5$, +6, and +7 \kms\ at bottom. These figures may
     be compared to the CNM structures in the more extended fields in
     Figs. \ref{Fig_Gauss_0}. The circle has a radius of 15\deg,
     corresponding to the central unapodized field of view. }
   \label{Fig_CNM_multi}
\end{figure*}

For each USM structure we determine the linewidth and the corresponding
Doppler temperature $T_{\mathrm D}$ as described by
\citet{Kalberla2016}. Alternatively, we use the CNM Gaussian parameters
from \citet{Kalberla2019} to calculate $T_{\mathrm D}$, which 
is an upper limit to the exitation or spin temperature of the \hi.  In
both cases $T_{\mathrm D}$ has, as expected for a turbulent medium
\citep{Vazquez1994}, a log-normal distribution 
\citep[e.g., ][Fig. 13]{Kalberla2016} and we determine independently for
each channel of the USM and CNM data the characteristic harmonic mean Doppler
temperature, see  Fig. \ref{Fig_Doppler} for the results.

Even though the selection criteria and numerical methods are
very different for the determination of USM and CNM structures, we find
 good agreement for both independent $T_{\mathrm D}$ measures. A
visual inspection of the data cubes confirms that USM filaments are
associated with the CNM, these filaments are always  located within CNM
structures. USM filaments with Doppler temperatures belonging to the
LNM or WNM temperature regimes have never been observed, though unsharp
masking was extensively used for quality control of survey data
observed with the Effelsberg and Parkes telescopes.

A comparison between Figs. \ref{Fig_Doppler} and \ref{Fig_Multi_Power}
shows that the local minima in $\gamma$, corresponding to steeper
spectral indices, are associated with cold \hi\ filaments. The
$T_{\mathrm D}$ curve for the USM data shows bumps that coincide with
the local minima in $\gamma$. The agreement is less good for the CNM
$T_{\mathrm D}$ curve.  We conclude that cold \hi\ small-scale
  structures are responsible for the observed steepening of thin slice
  spectral indices. As discussed before, these structures dominate the
  power spectra at low velocities.  Cold small-scale structures may be
considered as markers for phase transitions characterized by rapidly
reducing kinetic energy with decreasing physical size scale
\citep{Wareing2019}. Doppler temperatures $T_{\mathrm D} \la 200$ K for
the region discussed here are exceptionally low. The all sky HI4PI median
temperature is $T_{\mathrm D} \sim 223$ K \citep{Kalberla2016} and
$T_{\mathrm D} \sim 220$ K was found for the Arecibo sky
\citep{Clark2014}. An inspection of \citet[][Figs. 9, 10, and
  13]{Kalberla2018} indicates that there is cold gas in this region.
\citet{Peek2019} used \ion{Na}{i} absorption at high Galactic latitudes
to demonstrate that small-scale structures are colder than the
environment.

Our results support \citet{Clark2019}, and are in clear conflict with
\citet{Yuen2019} who deny the existence of cold \hi\ gas in the observed
region. This section further supports  the conjecture  that a steepening
of the spectral indices is found for particular cold \hi\ gas
condensations on small scales.

\section{CNM related to multiphase \hi }
\label{CNM}

The CNM is part of the multiphase \hi\ and there appears to be some
consensus that these CNM structures are restricted to small scale
structures that are embedded in the multiphase \hi\ (e.g. \citet{MO77}
and \citet{Wolfire2003}). Recent all-sky surveys, like the combined
Effelsberg and Parkes \hi\ survey \citep[HI4PI,][]{Winkel2016c} or the
Galactic Arecibo L-Band Feed Array Survey \citep[GALFA-\hi,][]{Peek2018}
have shown that many or most of the CNM structures need to be considered
as part of extended large-scale filamentary structures that are
associated with polarized dust emission \citep{Clark2019b}.

Fig. \ref{Fig_Multi_Power} rises the question what kind of
\hi\ structures might be responsible for the spectral index changes at
velocities around $v_{\mathrm{LSR}} = -6$, 1, and 6 \kms. The average
\hi\ emission across the the field of view is smooth and without any
structures that could resemble fluctuations in
Fig. \ref{Fig_Multi_Power} \citep[see][Fig. 1]{Clark2019}. Power spectra
are however affected by emission fluctuations rather than by the average
emission and it was shown already in early \hi\ absorption line studies
that rms emission fluctuations are related to the line widths of the
corresponding CNM absorption lines (\citet{Lazareff1975} and
\citet{Mebold1982}). These authors proposed an explanation in terms of
turbulence. \citet{Mebold1982} interpret these features as arising in
predominantly atomic neutral shells around molecular hydrogen and dust
clouds.

In Sect. \ref{Cold} we discussed the proposal that steep power spectra
are related to low Doppler temperatures. The implication would be that
velocity channels with low $\gamma$ values are dominated by
\hi\ emission from cold small scale structures with a lot of
fluctuations on all scales. To verify this proposal we compare in
Fig. \ref{Fig_CNM_multi} the total multiphase \hi\ emission with the CNM
emission as derived from our Gaussian analysis. For a better
visualization of weak small scale structures we zoom in to the central
unapodized region of our field of view, we clip also a part of the
strongest multiphase emission structures. Fig. \ref{Fig_CNM_multi} shows
on top channel maps at $v_{\mathrm{LSR}} = -15$, +15, and -6 \kms, below
at 5, 6, and 7 \kms. In case of $v_{\mathrm{LSR}} = -15$ \kms\ the
multiphase emission is mostly diffuse but contains some CNM
structures. For $v_{\mathrm{LSR}} = +15$ \kms\ the emission is in
general very weak. The case $v_{\mathrm{LSR}} = -6$ \kms\ is very
different with strong emission for multiphase \hi\ as well CNM over most
of the field of view. These structures are very complex and require
detailed investigations using different intensity windows for different
parts of the field. We investigated also the channel map at
$v_{\mathrm{LSR}} = 1$ \kms, the situations is there even more
complex. We don't show these data and skip also a detailed discussion of
the $v_{\mathrm{LSR}} = -6$ \kms\ to avoid an unnecessary lengthy
presentation. At bottom of Fig. \ref{Fig_CNM_multi} we present data for
$v_{\mathrm{LSR}} = 5$, 6, and 7 \kms.  These figures demonstrate two
effects. In the first instance the emission structures change
systematically but significantly when incrementing the velocity by only
1 \kms. It is also obvious from these less crowded fields that most of
the CNM structures trace prominent filamentary multiphase
\hi\ structures. Many of the CNM structures persist only for three
channels, implying a low velocity width or Doppler temperature. Such a
FWHM velocity width of 3 \kms\ corresponds to a Doppler temperature of
$T_{\mathrm D} \sim 200$ K, a value that was found previously
characteristic for filamentary structures (e.g. \citet{Clark2014},
\citet{Kalberla2016}, and \citet{Kalberla2020}).  Several prominent
filaments show up with parallel structures for a length of 10\deg to
20\deg, resulting in wavy appearance.

\begin{figure}[th] 
    \centering \includegraphics[width=9cm]{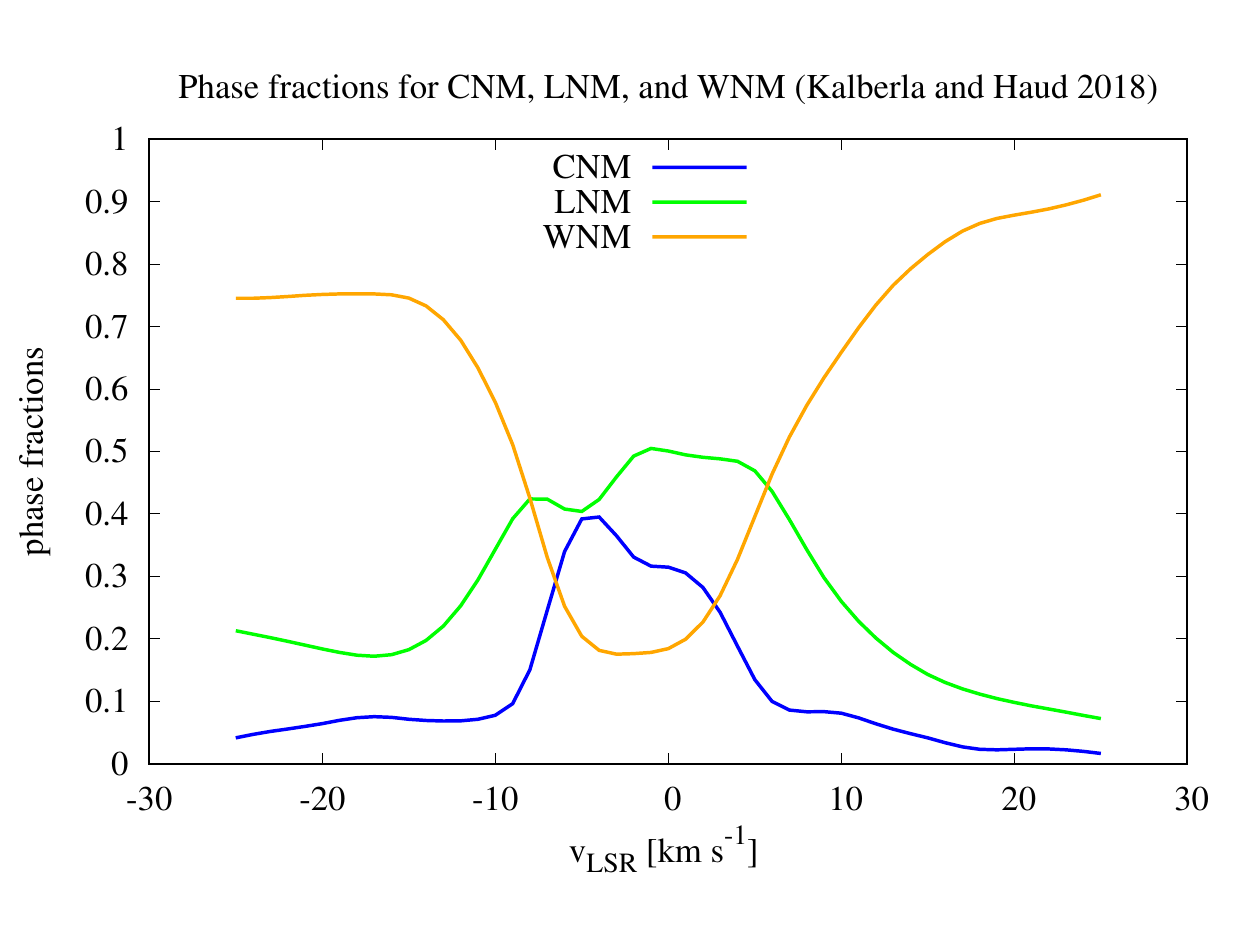}
    \caption{Velocity distribution of average \hi\ phase fractions for
      CNM, LNM, and WNM within the inner unapodized field of view with a
      radius of 15\deg. }
   \label{Fig_phase_fract}
\end{figure}

Interpreting structures in Fig. \ref{Fig_CNM_multi} we need to take
selection criteria into account. Only a fraction of the data can be
presented and we have chosen two channels at $v_{\mathrm{LSR}} = -15$
and +15 \kms\ as representative for the wings of the $\gamma$
distribution from Fig. \ref{Fig_Multi_Power}. This selection results in
multiphase \hi\ structures that contain only a low fraction of CNM with
typical line widths of $\Delta v_{\mathrm{LSR}} \sim 4 $ \kms. For
velocities $-10 \la v_{\mathrm{LSR}} \la 8$ \kms\ we observe a
completely different picture. Except for a diffuse background with
$T_\mathrm{B} \sim 2$ K the \hi\ distribution displayed in the bottom
panels of Fig. \ref{Fig_CNM_multi} is dominated by strong CNM emission
with low Doppler temperatures. The visual impression of PPV data cubes
may be subjective and can be biased by the way the data are
presented. We based our selection on results from statistical methods
that led to Figs. \ref{Fig_Multi_Power} and \ref{Fig_Doppler}. These
results are objective and unbiased. One may question a relation between
spectral index $\gamma$ (Fig. \ref{Fig_Multi_Power}) and harmonic mean
Doppler temperature (Fig. \ref{Fig_Doppler}) but the visual inspection
of Fig. \ref{Fig_CNM_multi} supports an interpretation that steep power
spectra are caused by particular strong filamentary \hi\ structures that
contain significant amounts of CNM with low Doppler temperatures. Cold
filamentary or wavy \hi\ structures were previously found to be
associated with anisotropies in the power distribution and steep
spectral indices (\citet{Kalberla2016b} and \citet{Kalberla2017}) and
these agreements can not be dismissed as purely accidental. It was also
shown previously that spectral indices for the CNM depend on the CNM
phase fractions \citep{Kalberla2019}. To verify such dependencies we
determine average CNM, LNM, and WNM phase fractions as defined by
\citep{Kalberla2018} within the inner unapodized field of view with a
radius of 15\deg. These phase fractions are shown in Fig.
\ref{Fig_phase_fract} for comparison with Figs. \ref{Fig_Multi_Power}
and \ref{Fig_Doppler}. Within the velocity range $-8 \la
v_{\mathrm{LSR}} \la 6$ \kms\ most of the the multiphase \hi\ is cold.


We have chosen for the display in Fig. \ref{Fig_CNM_multi} a lower limit
$T_{\mathrm B} = 0.3$ K that was used for Gaussian emission structures
throughout this analysis. This value corresponds to an approximate three
sigma peak noise level; the rms uncertainties for EBHIS amount to 90 mK
\citep{Winkel2016c}. The analysis of our Gaussian decomposition
algorithm \citep{Haud2000} with the LDS data, which have similar rms
uncertainties of about 80 mK \citep{Hartmann1997}, has demonstrated that
for Gaussians with larger central heights, $T_{\mathrm B0}$ , our
results are in general agreement with the independent decompositions by
other authors. The median height of the unmatching Gaussians from
different decompositions of the identical profiles was 0.1 K. We
analyzed also the dependence of the obtained Gaussians on the
observational noise and found that such influence is strongest for
Gaussians with $T_{\mathrm B0} < 0.3$ K. From this we conclude that the
Gaussian parameters for the CNM with $T_{\mathrm B0} \ga 0.3$ K are
statistically significant. A general feature of Fig.
\ref{Fig_CNM_multi} for all velocities is that small scale CNM
structures are in any case embedded in more diffuse multiphase
\hi\ structures, as expected from theory (\citet{MO77} and
\citet{Wolfire2003}).

\begin{figure*}[th] 
   \centering
   \includegraphics[width=5.2cm,angle=-90]{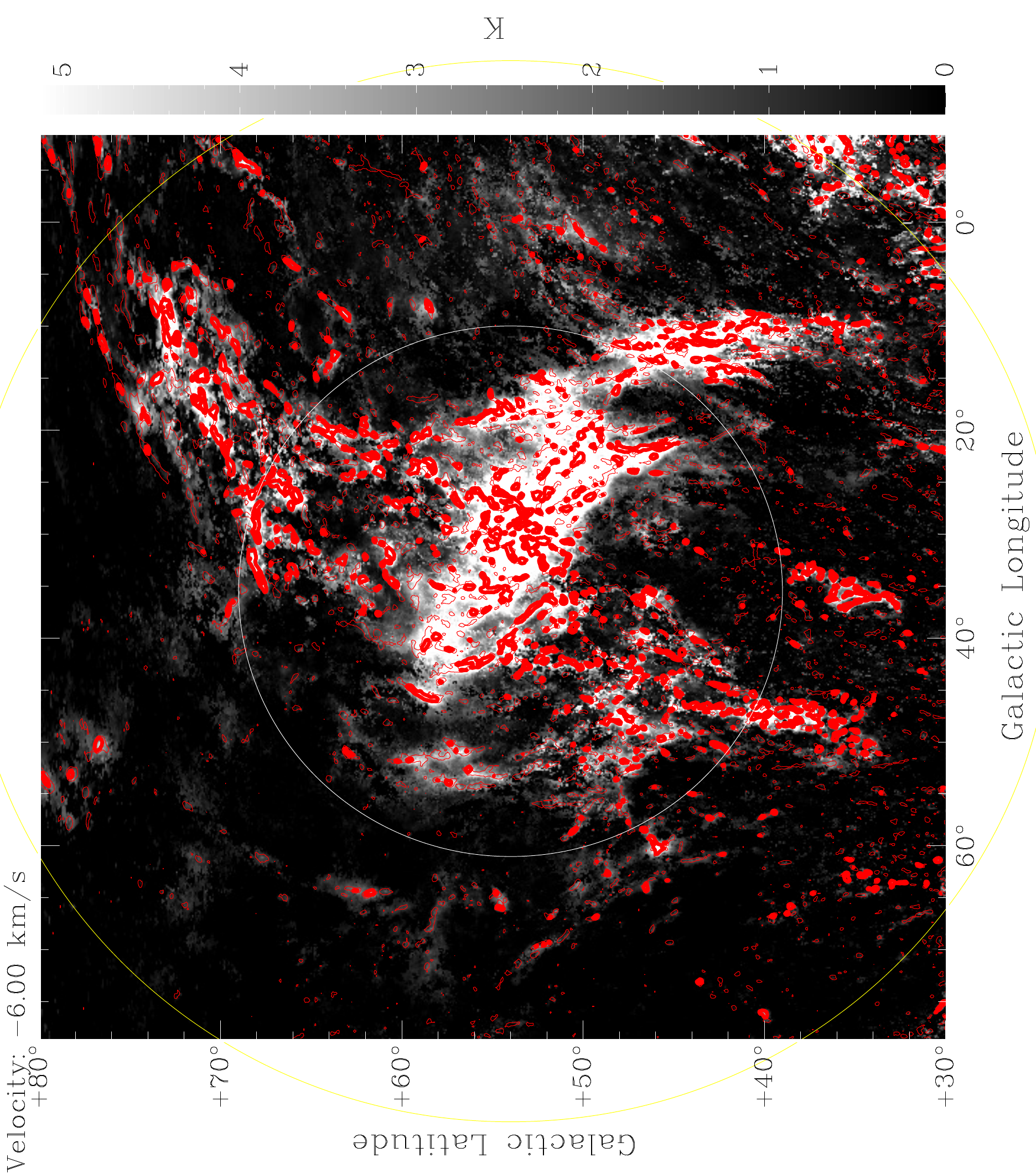}
   \includegraphics[width=5.2cm,angle=-90]{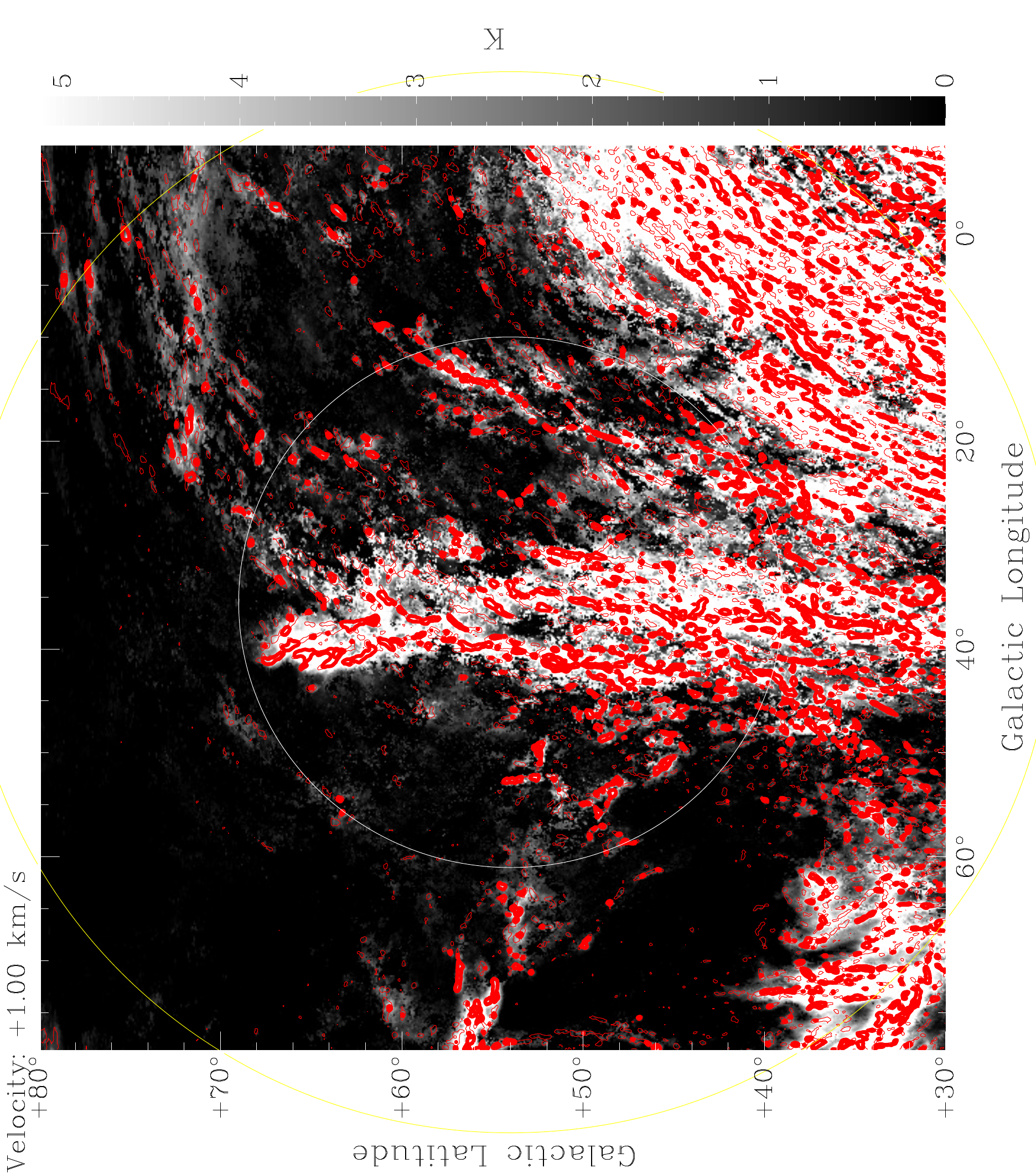}
   \includegraphics[width=5.2cm,angle=-90]{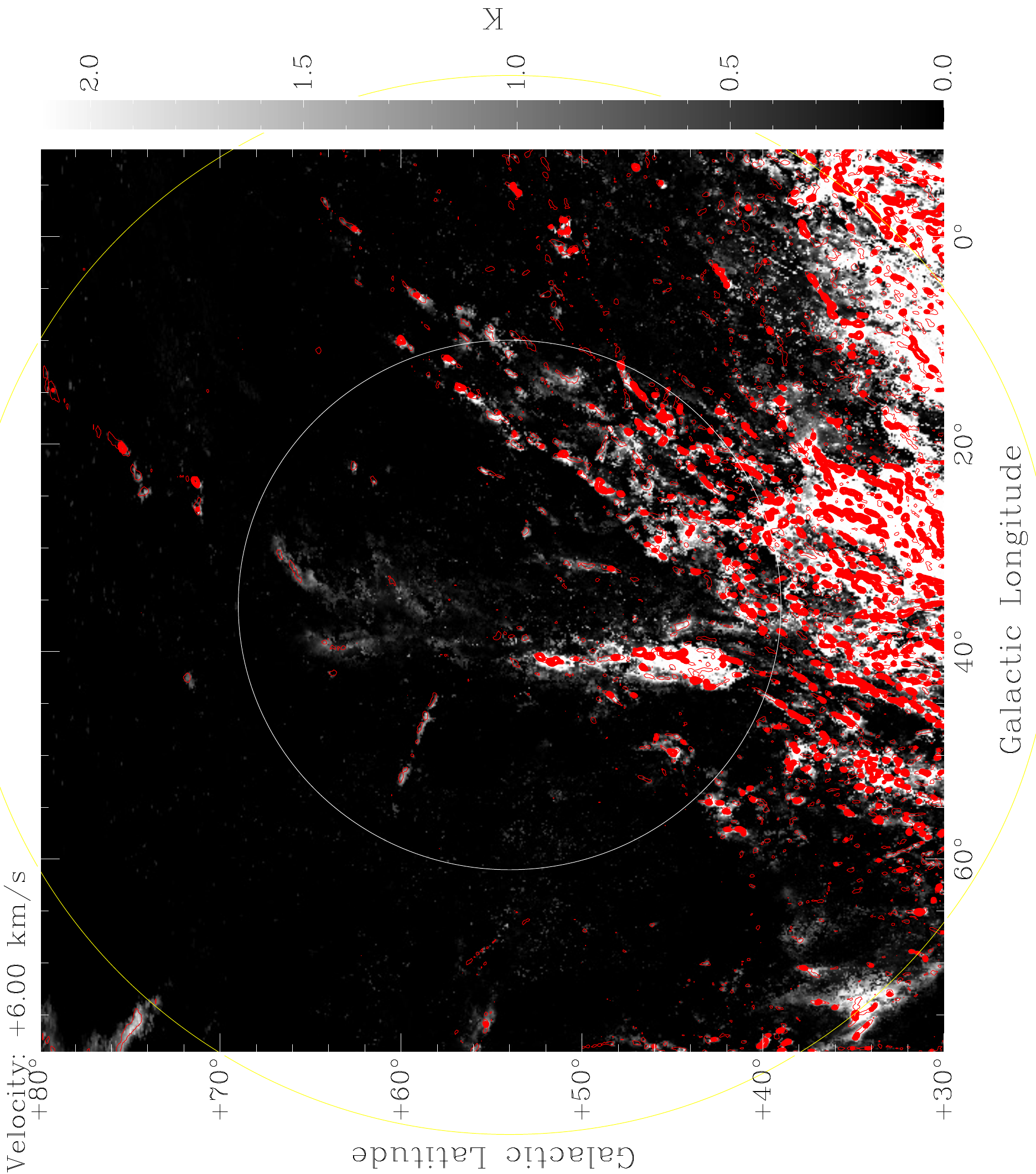}
   \includegraphics[width=5.2cm,angle=-90]{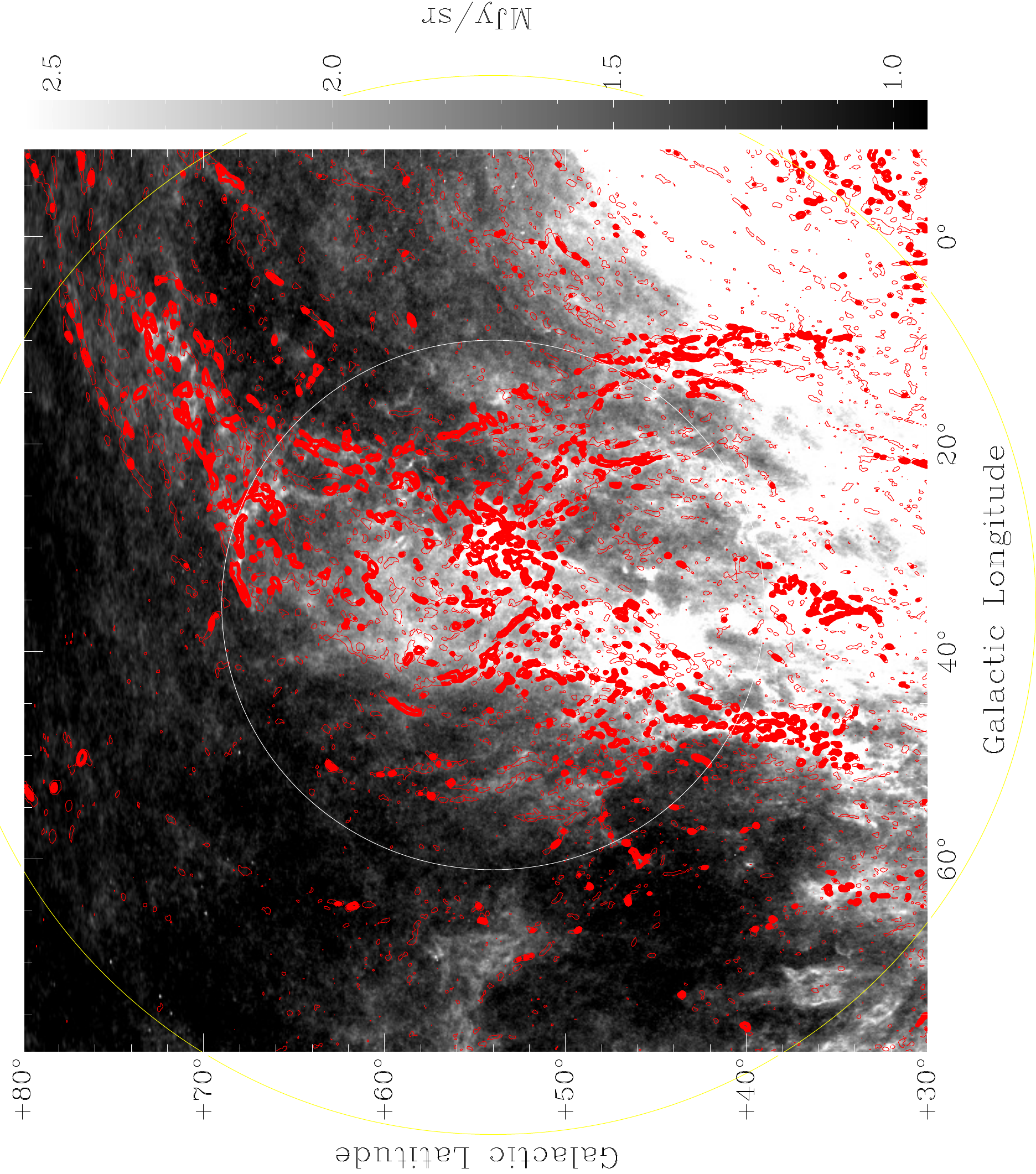}
   \includegraphics[width=5.2cm,angle=-90]{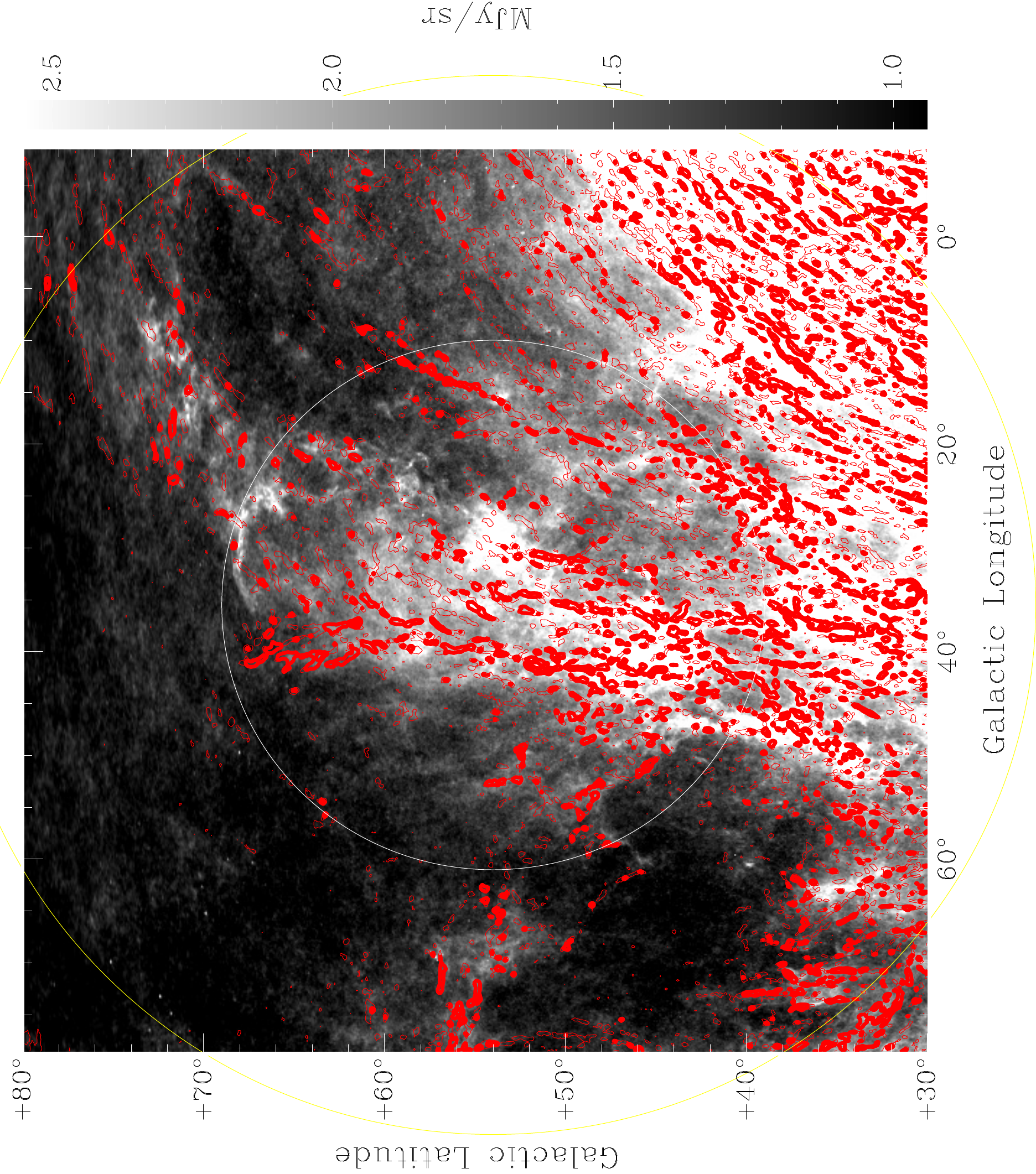}
   \includegraphics[width=5.2cm,angle=-90]{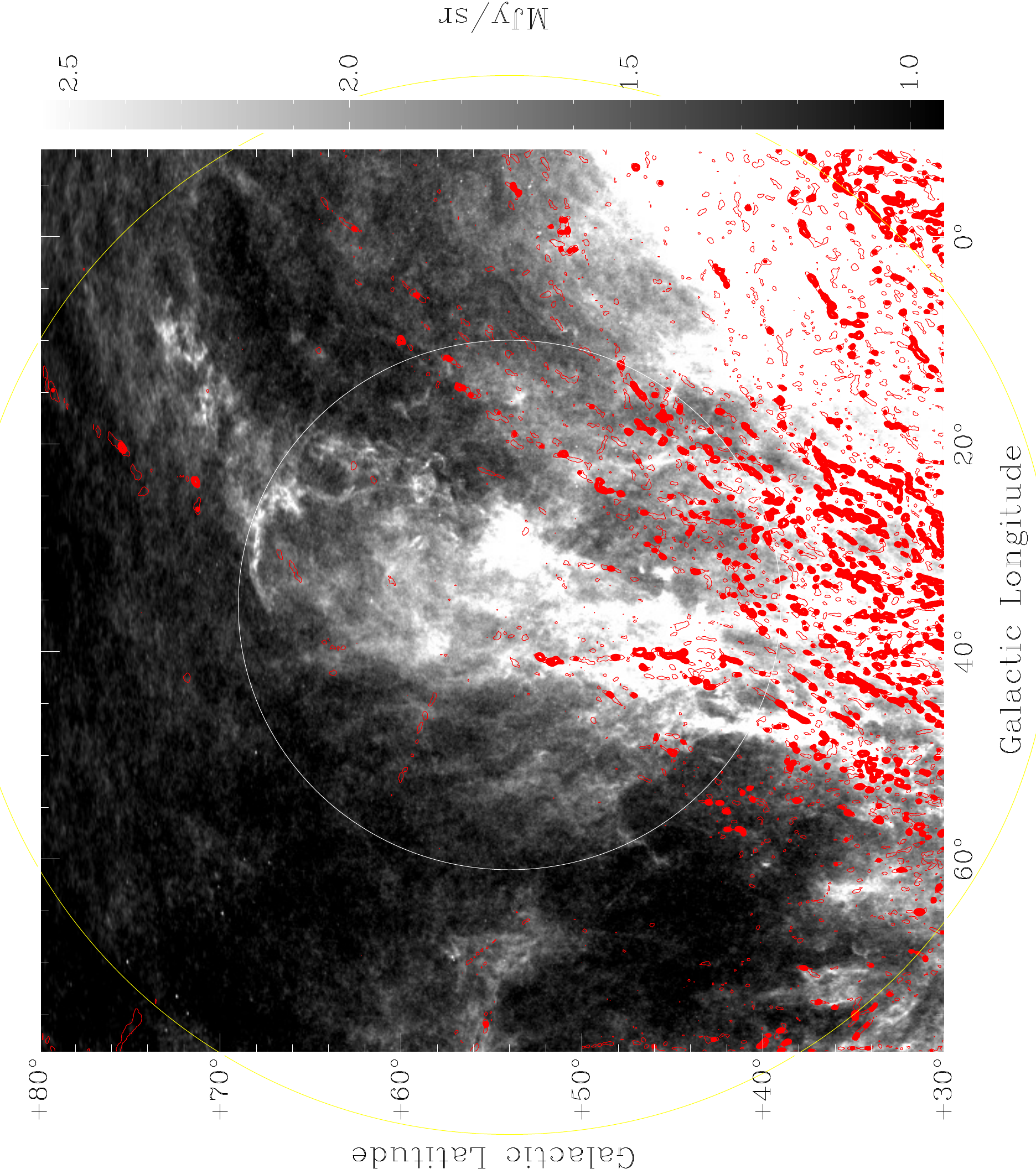}
   \caption{Upper plots: CNM emission at velocities $v_{\mathrm{LSR}} = -6,
     +1$, and $+6$ \kms\ (from left to right), overlayed with red contours
     from USM emission at the same velocities with a level of $>0.3$
     K. The lower plots show FIR emission at 857 GHz as observed by {\it
       Planck} (left), overlayed with the same USM structures as on
     top. The circles have radii of 15\deg\ (white) and 30\deg\
     (yellow). }
   \label{Fig_Gauss_0}
\end{figure*}

\section{The relation between dust and cold
  \hi\ gas}
\label{Visual}

The third key question that we like to solve concerns the relation
between dust and cold gas. The {\it Planck} 857 GHz dust emission is
dominated by thermal dust and its intensity is proportional to dust
column density, which is expected to follow the \hi\ density
\citep{Yuen2019}. There appears to be a general consensus that
\hi\ gas and dust are frequently filamentary and the FIR emission is
closely aligned with prominent USM structures. The debate is whether
these filaments indicate cold and dense gas or velocity caustics in a
turbulent medium, independent of density and temperature or line
width.  In the previous section we  show that USM structures are
on average as cold as the CNM with Doppler temperatures even below 200
K (Fig. \ref{Fig_Doppler}). Here we  explore the spatial
relations between FIR, USM, and CNM structures. \citet{Clark2019}
demonstrated an  increase in the $I_{857} / N_{HI}$ ratio at positions
with increasingly stronger USM emission. Such a relation was
questioned by \citet{Yuen2019}.

The power spectra in Fig. \ref{Fig_Multi_Power} have three well-defined local minima. A visual inspection of the PPV cubes of the USM
and CNM emission shows that these extrema are caused by the
\hi\ source distribution. We observe three major cloud complexes at
velocities $v_{\mathrm{LSR}} \sim -6, 0$, and $+6$ \kms. Depending on
velocity, the USM structures are embedded in different CNM
structures. This is demonstrated in the upper panels of
Fig. \ref{Fig_Gauss_0}. The USM structures always follows the CNM. The
lower panels show the relation between USM and FIR emission. FIR
filaments are associated with distinct USM structures at different
velocities and  all of the USM channel maps need to be inspected to
identify counterparts between FIR and USM data. The \hi\ distribution
is very complex, but on visual inspection all of the USM structures at
different velocities appear to be related to FIR emission.

\begin{figure*}[th] 
   \centering
   \includegraphics[width=5.2cm,angle=-90]{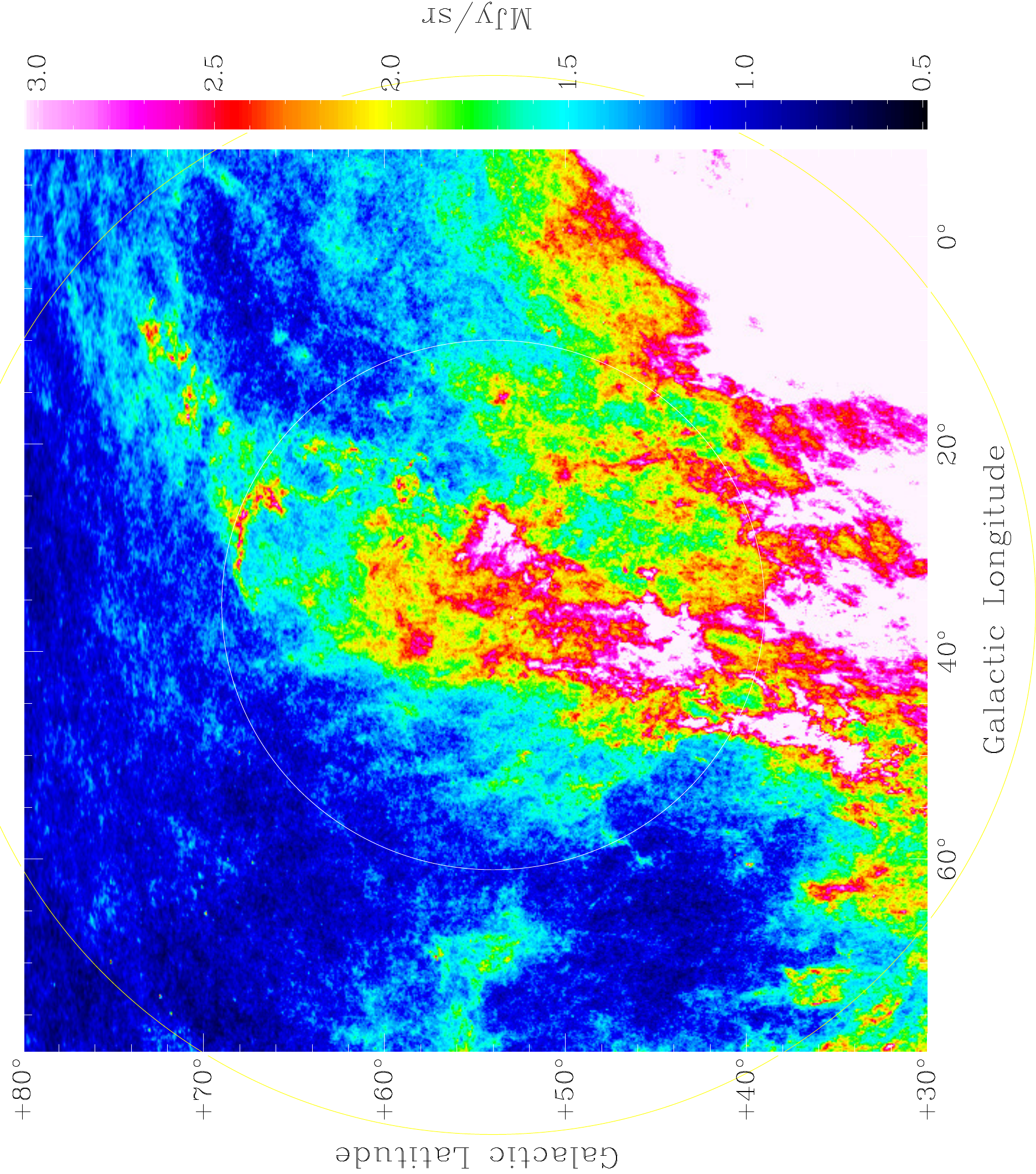}
   \includegraphics[width=5.2cm,angle=-90]{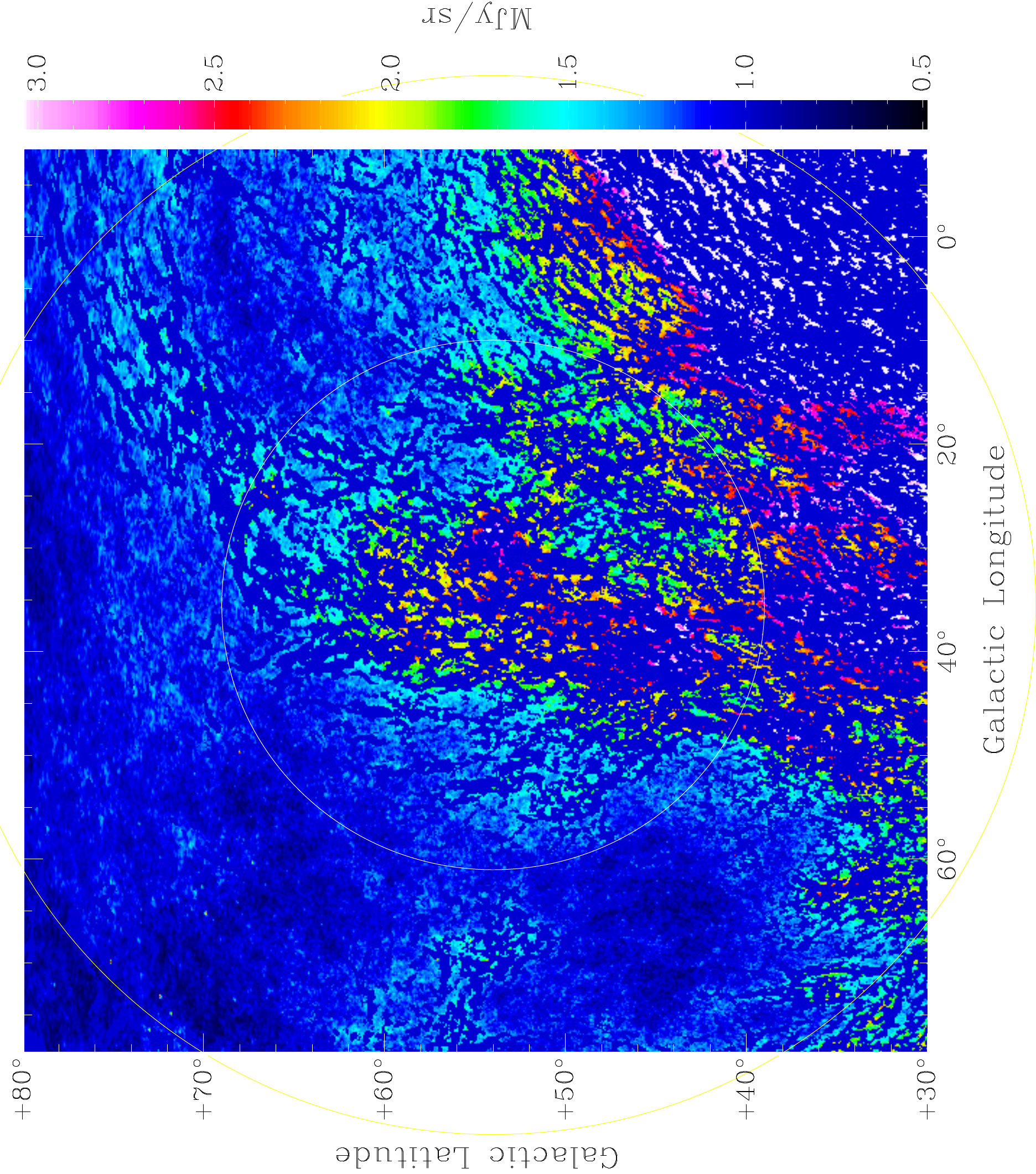}
   \includegraphics[width=5.2cm,angle=-90]{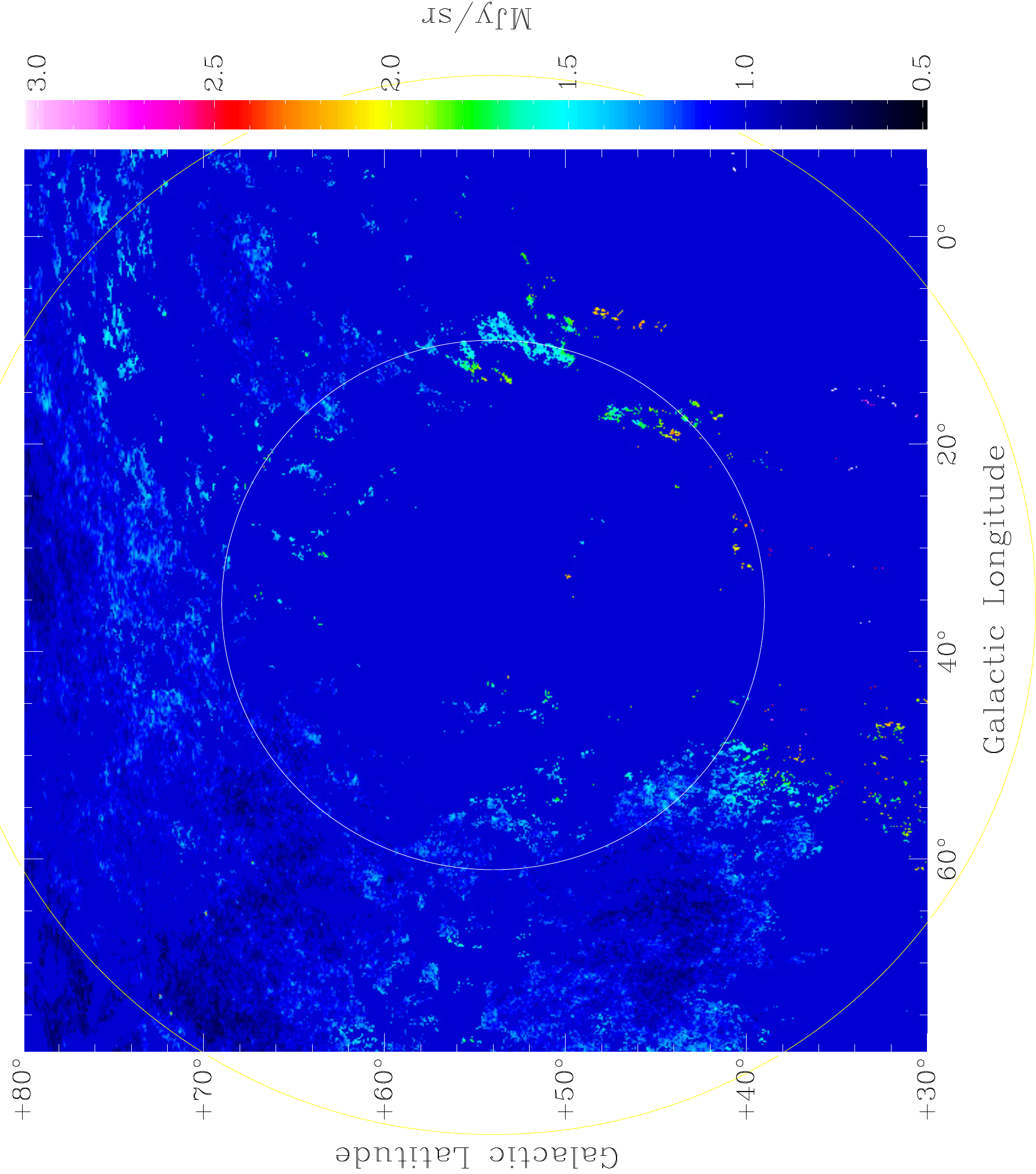}
   \caption{Far-IR emission at 857 GHz as observed by {\it Planck} (left),
     masked in presence of USM structures (middle) and CNM emission
     (right) at levels of $>3$ K. The circles have radii of 15\deg\
     (white) and 30\deg\ (yellow).}
   \label{Fig_HFI_USM}
\end{figure*}

For a further numerical analysis of the relation between FIR and
\hi\ emission we adopt the hypothesis that all USM structures at all
velocities between $ -25 < v_{\mathrm{LSR}} < 25 $ \kms\ are
associated with dust, hence FIR emission at 857 GHz. We flag the FIR
map at all positions with USM emission with brightness temperatures $>
0.3$ K and replace the observed FIR emission with a constant
background level. Following \citet{Planck2016b} we determined this
background level to $1.06 \pm 0.16 $ MJy sr$^{-1}$ within a $12\fdg5$
x $12\fdg5$ field at (GAL$ = 90\deg$, GAB$ = -80\deg$) as a 857 GHz
FIR background that is unaffected by USM or CNM features; we did not
apply any offset corrections for warmer \hi\ components.  Figure
\ref{Fig_HFI_USM} shows the results after flagging. In  the left panel we
display for comparison the observed 857 GHz FIR emission, in the
middle the FIR after flagging for dust associated with USM structures.
After flagging the FIR map for USM signals we repeat the calculations
for CNM emission. In the right panel we display
the results after flagging for CNM counterparts.

The middle plot of Fig. \ref{Fig_HFI_USM} shows that 83.6\% of the FIR
emission has been flagged under the assumption that USM and FIR
emission are spatially associated. The remaining FIR flux originates
close to USM structures, implying that the dust distribution is
slightly more diffuse than the USM emission. Inspecting the relation
between USM and CNM structures we find that the CNM is in general
somewhat more extended than the USM filaments (see
Fig. \ref{Fig_Gauss_0} top). FIR emission can  be closley linked to the
CNM. The result from such a hypothesis is plotted in the right panel of Fig. \ref{Fig_HFI_USM}.  We find that now 99.6\% of the FIR
has been flagged. This ratio could  be increased by selecting a
velocity range of $ -30 < v_{\mathrm{LSR}} < 25 $ \kms. After a visual
inspection of the data we found no indication of dust associated
with CNM outside this range.  The remaining unflagged positions in
Fig. \ref{Fig_HFI_USM} (right; 20\% of the observed field)
have an average FIR flux density of $1.07 \pm 0.18 $ MJy sr$^{-1}$,
consistent with our previous assumption of a $1.06$ MJy sr$^{-1}$
background level. We conclude that a significant fraction of the FIR
emission originates from dust that is associated with the CNM. USM
structures are spatially embedded within the CNM (see also
Fig. \ref{Fig_Gauss_0}) and associated with a major part (83.6\%) of
the dust.

After demonstrating the spatial association between dust and USM
structures we want to clarify whether the ratio $I_{857} / N_{HI}$ 
depends on the strength of the USM intensity, as observed by
\citet{Clark2019}. These authors used only a fraction of the available
data and restricted their analysis to the main \hi\ component close to
zero \kms. Their results were questioned by \citet{Yuen2019}, but these
authors used  only a small 9\deg\ x 9\deg\ subset of the database. Here we
 analyze all velocities $-25 < v_{\mathrm{LSR}} < 25 $ \kms\ and
the whole field of view. As shown by Fig. \ref{Fig_Doppler}, the cold
gas is concentrated around low velocities; therefore, we integrate the
column densities $N_{HI}$ only for velocities in the range $-25 <
v_{\mathrm{LSR}} < 25 $ \kms. For each position we search for USM
emission at all velocities. When  we find several USM components
at different velocities along the line of sight we assume that these
components have the same $I_{857} / N_{HI}$ value. This is an ambiguity
in our analysis, but the results do not change if we 
disregard such positions.


\begin{figure}[th] 
    \centering
    \includegraphics[width=9cm]{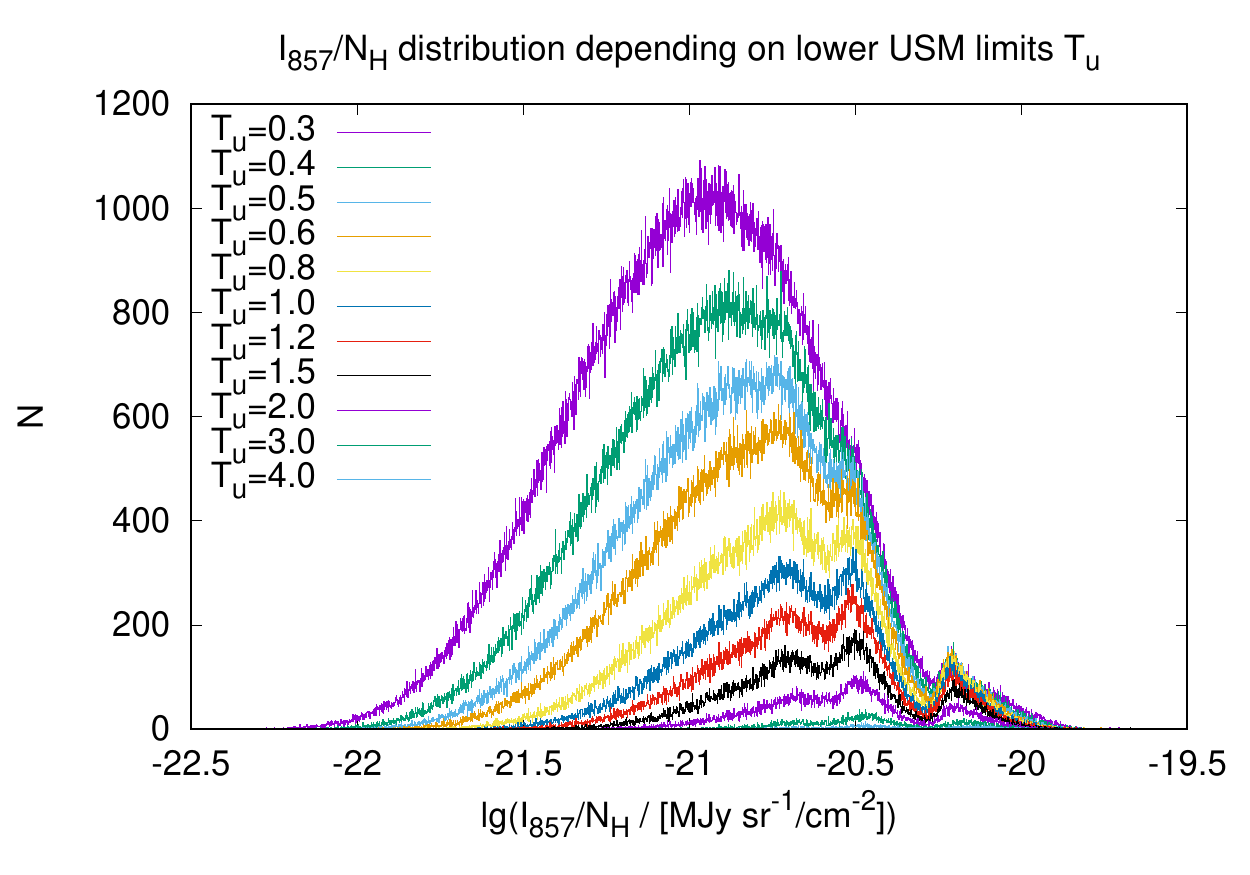}
    \caption{Histograms for the $I_{857}/N_{H}$ distributions as
      a function of the lower limits $T_{u}$ for the USM emission used
      to select the data.}
   \label{Fig_Plot_X_hist_USM}
\end{figure}

\begin{figure}[th] 
    \centering
    \includegraphics[width=9cm]{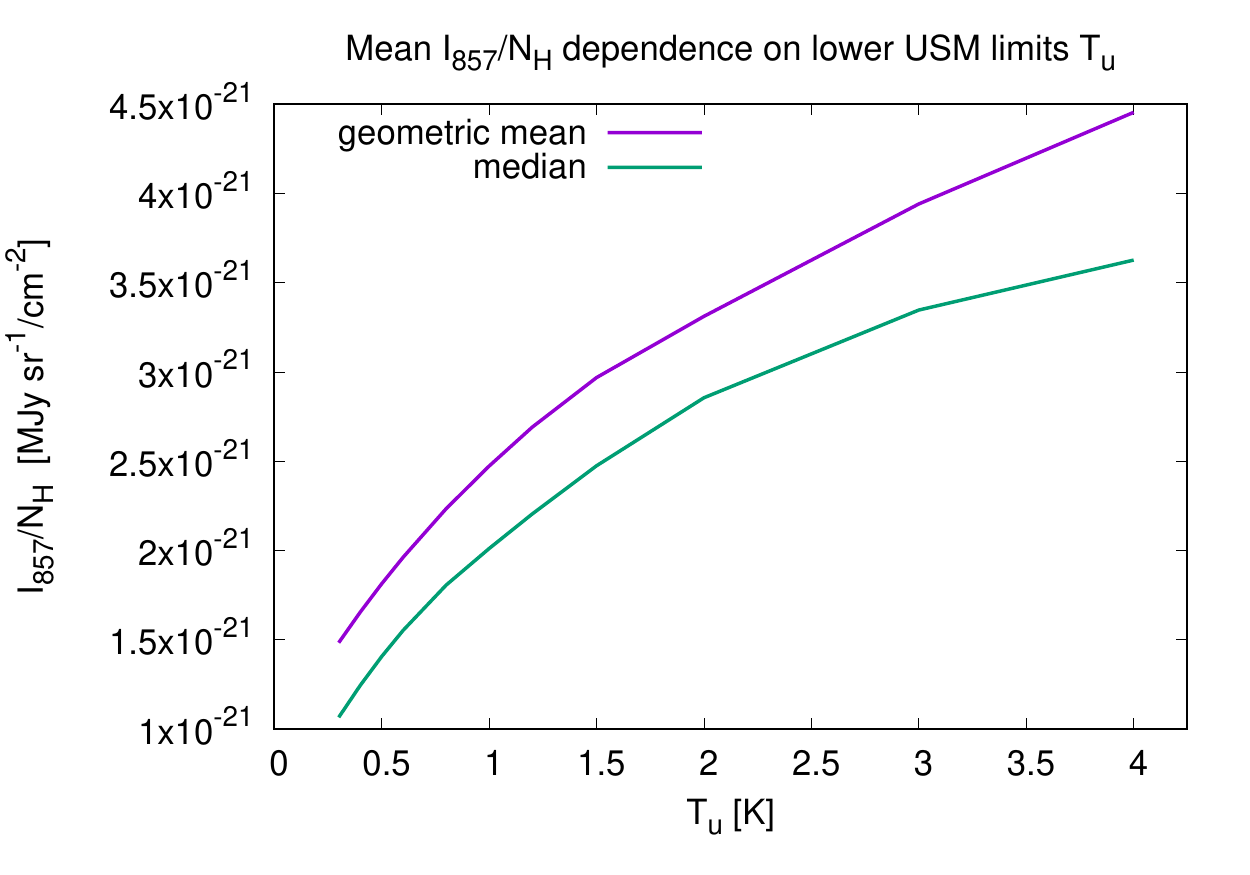}
    \caption{Dependence of the $I_{857}/N_{H}$ averages on the lower
      USM limits $T_{u}$ used for Fig. \ref{Fig_Plot_X_hist_USM};
      shown are geometric mean and median. }
   \label{Fig_Plot_X_hist_USM_drift}
\end{figure}

Figure \ref{Fig_Plot_X_hist_USM} shows histograms for the
$I_{857}/N_{HI}$ distributions depending on lower limits $T_u$ for USM
intensities. We find that for increasing $T_u$ (thus for more intense
USM signals) the $I_{857}/N_{HI}$ distribution shifts to higher
values. At the same time we find a significant decrease in the total
number of such features. The $I_{857}/N_{HI}$ distribution for $T_u=0.3$
(all significant USM structures selected) is, as expected for a
turbulent medium, approximately log-normal, but this behavior breaks down
for higher $T_u$ limits. We calculate  the geometric mean and the median for the derived $I_{857}/N_{HI}$
distributions. Figure
\ref{Fig_Plot_X_hist_USM_drift} shows that the mean value for
$I_{857}/N_{HI}$   increases continuously with increasing $T_u$
limit. This increase is significant; our results support the findings by
\citet{Clark2019}; however,   as stated by \citet{Yuen2019},  the
increase in   $I_{857}/N_{HI}$  for more pronounced USM
structures may be violated in some smaller specific sky patches. Such cases
cannot be used as an argument against global and well-defined trends in
a turbulent medium.


\begin{figure}[th] 
    \centering
    \includegraphics[width=8cm,angle=-90]{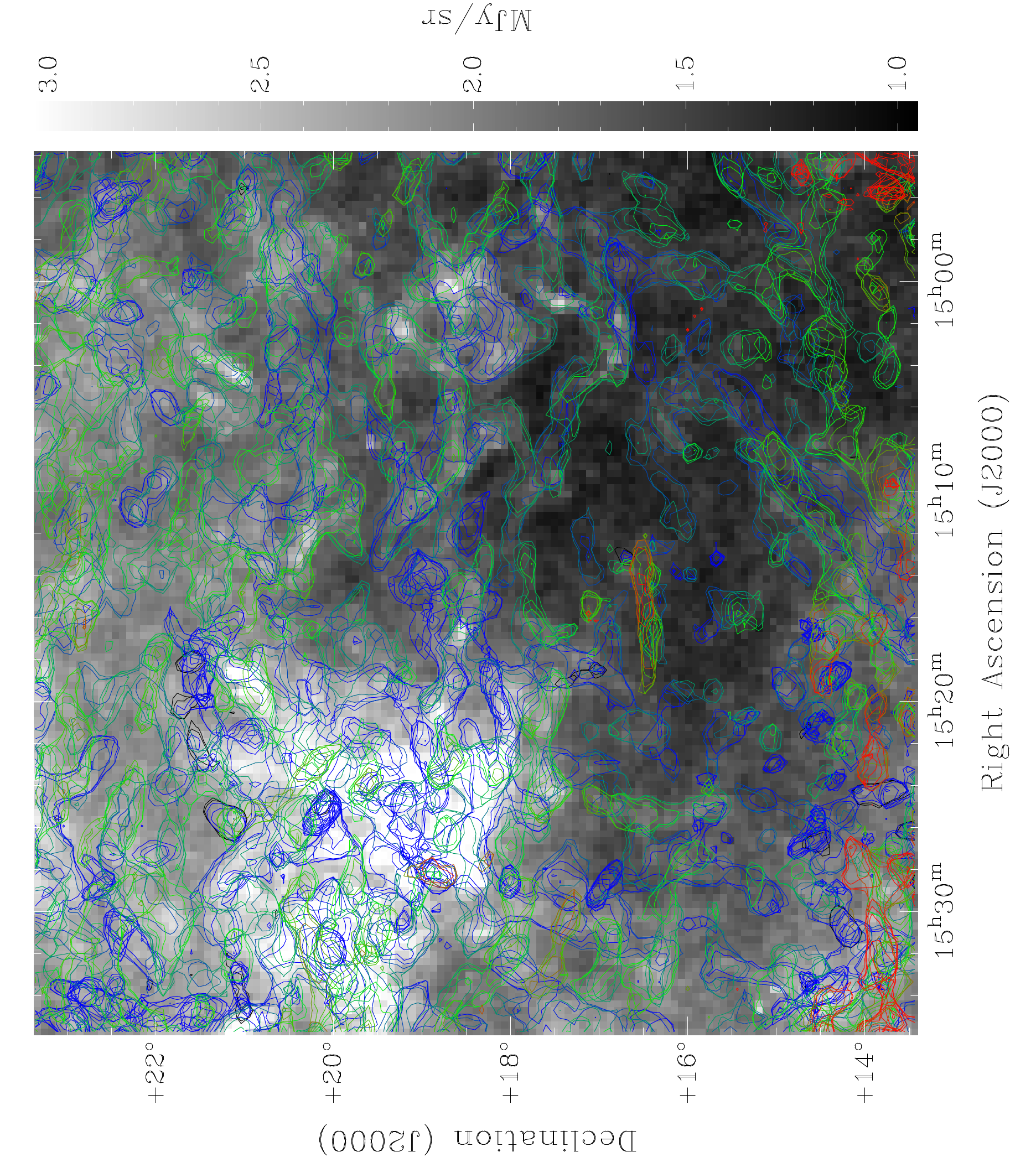}
    \caption{Renzograms of USM structures in the field discussed by
      \citet{Yuen2019} at a contour level of 0.3 K; colors represent the
      velocity field of the USM structures, from negative (blue) 
      to positive (red) radial velocities. The gray scale image
      in the background shows the FIR emission at 857 GHz. }
   \label{Fig_Plot_Power_Yuen_F3_Vel}
\end{figure}

The \hi\ distribution and in particular the USM structures can be very
complex, making   the interpretation difficult. Figure
\ref{Fig_Plot_Power_Yuen_F3_Vel} shows Renzograms for the field used
by \citet{Yuen2019} as the basis for counter-arguments against
\citet{Clark2019}. Plotted are contour levels for USM intensities of
0.3 K, overplotted for each velocity channel to the FIR distribution
in the background.  We see here a terribly complicated case with
numerous USM worm structures, much more complex than the  filamentary
USM structures that were considered by \citet[][Fig. 8]{Clark2019}.

\section{Discussion}
\label{Discussion}

In Sections \ref{Power} to \ref{Visual} we discussed key questions that
emerged from the dispute between \citet{Clark2019} and \citet{Yuen2019}
about the nature of observed \hi\ structures. We analyzed the same
region on the sky as discussed by these authors, but used completely
independent data from the EBHIS survey \citep{Winkel2016a}. We  also used
different analysis tools; our intention was to verify the results under
discussion. In this section we aim to broaden the discussion, taking 
particular results from the literature into account that have been neglected
so far.


It has been noted by \citet{Yuen2019} that the most valuable insight
from \citet{Lazarian2000} is the prediction of the spectral slope change
between the thin and thick PPV slices that is uniquely related to the
spectral indices of turbulent velocity and density. A clear support to
the VCA predictions was given by \citet{Stanimirovic2001}, but on the
other hand several authors found it difficult to apply VCA for an
unambiguous separation of spectral indices between turbulent velocity
and density in the Milky Way (
\citet{Deshpande2000}, \citet{Dickey2001},
\citet{Miville2003a,Miville2003b}, \citet{Khalil2006}, \citet{Roy2010},
\citet{Pingel2018}, \citet{Kalberla2016b}, \citet{Kalberla2017},
\citet{Blagrave2017}, \citet{Choudhuri2018}, and \citet{Kalberla2019}.
Many of these authors report that the steepening is restricted to a
narrower range in velocity width ($6 \la \Delta v_{\mathrm{LSR}} \la 17
$ \kms) than expected ($\Delta v_{\mathrm{LSR}} \ga 17 $ \kms), similar
to our finding here. The conclusion is often that there is no
significant change in the exponent with the velocity width of the slice.
However \citet[][Table 3]{Yuen2019} redetermine spectral indices for
density and velocity and in a few cases draw completely different
conclusions from the authors in the original publications of
\citet{Deshpande2000}, \citet{Khalil2006}, and
\citet{Choudhuri2018}. Unfortunately neither methods nor reasons for the
updates were explained by \citet{Yuen2019}, so we need to wait for a
further publication of details that can illuminate these
discrepancies. It is mandatory to explain what led to such changes. As
long as clear statements are missing, the new results in Table 3 of
\citet{Yuen2019} remain highly questionable.

A one-to-one relation of changes in spectral slope between thin and
thick PPV slices, as predicted by VCA, is hampered by velocity
dependences of thin slice spectral indices, as observed by
\citet{Kalberla2016b}, \citet{Kalberla2017}, \citet{Blagrave2017},
\citet{Choudhuri2018}, and \citet{Kalberla2019}.  A common finding of
these authors is that \hi\ structures in channel maps can change
dramatically over even small changes in velocity.  Rapid intensity
changes with velocity imply in any case a dominance of cold (narrow
velocity width) structures. Such changes also affect the power spectra;
thin slice spectral indices depend on thermal conditions.  Power spectra
for the multiphase \hi\ gas steepen as the CNM contribution gets colder,
suggesting that the \hi\ phase composition is affecting the turbulent
state of the \hi. Simulations show that phase transitions cause steep
power spectra \citep{Wareing2019} since thermal instabilities
dynamically can form sheets and filaments on typical scales of 0.1–0.3
pc. These simulations show a growth of structures, associated with a
rapid rise and steepening of the density power spectra. Our conclusion
from Figs. \ref{Fig_Multi_Power} and \ref{Fig_Doppler}, consistent with
the findings of \citet{Kalberla2016b}, \citet{Kalberla2017}, and
\citet{Kalberla2019}, is that the steepest multiphase power spectra are
associated with particularly cold \hi\ gas, observed as CNM and USM
structures. Such thin slice power spectra show changes in spectral
indices that are far stronger than the observed VCA steepening of the
power spectra with velocity width, as proposed by \citet{Lazarian2000}
and \citet{Yuen2019}. Our results support the assertion by
\citet{Clark2019} that changes in the CNM content predominantly cause
the observed fluctuations in spectral indices.

Caustics are caused by geometrical effects; the velocity field can push
gas at the same velocity along the line of sight, mimicking density
enhancements. Other geometrical considerations may offer an
interpretation for the steepening of the power spectra. For optically
thin gas, \citet{Lazarian2000} predict a steepening as the spatial
thickness of the analyzed region increases. For a determination of the
3D power-law exponent of the underlying density field the thickness of
the sampled region should be at least as large as the dimension of the
map perpendicular to the line of sight. In general, it is difficult to
estimate the thickness of an observed \hi\ layer; however, applying
these considerations to the observed steepening of the \hi\ from
Fig. \ref{Fig_Multi_Power} would imply that the depth of the thin slice
multiphase \hi\ distribution, as well as the depth of the CNM, would
increase at the peak velocities for the spectral indices
($v_{\mathrm{LSR}} = -6 $, 0 +6 \kms). At the same time we observe that
this gas is particularly cold (Fig. \ref{Fig_Doppler}). A similar
situation holds for other sources discussed by \citet{Kalberla2016b} and
\citet{Kalberla2017}. The conclusion that the extension of \hi\ layers
along the line of sight must be inverse to their Doppler temperatures is
so strange that further discussion is not needed.

It was shown by \citet{Kalberla2019} that the multiphase \hi\ should be
considered as a phase composite with CNM, LNM, and WNM. The associated
individual power spectra are correlated and differ significantly from
the multiphase case. For a complete description it is necessary to take
also cross-correlations between these phases into account
\citep[][Eq. 4]{Kalberla2019}. All of the power spectra for the
individual phases are shallow in comparison to the multiphase case with
enhanced power for high multipoles. These are spatial frequencies where
USM filaments can be observed with large single-dish telescopes. Power
spectra for the individual phases at intermediate multipoles $ 10 \la l
\la 100 $ are more shallow and for the CNM $\gamma_{\mathrm{CNM}} \sim
-2.4$ are in far better agreement with $\gamma_{\mathrm{857}} \sim -2.5$
than the average multiphase index $\gamma_{\mathrm{NHI}} \sim
-2.9$. Such an agreement is expected if FIR emission is correlated with
CNM and USM structures as proposed by \citet{Clark2019}.  We show in
Sect. \ref{Visual} that the value of $I_{857}/N_{HI}$ increases with
increasing strength of the USM signal. The implication is that either
the dust emissivity or dust-to-gas ratio is changed in filaments or that
CNM filaments are associated with molecular hydrogen (\citet{Clark2019}
and \citet{Kalberla2020}). Optical depth effects have been ruled out by
\citet{Kalberla2019}.

Our analysis of the correlation between the FIR emission at 857 GHz and
\hi\ structures follows that of \citet{Clark2014} and
\citet{Kalberla2016}; we use small-scale structures from USM data for
correlations with FIR emission. \citet{Lazarian2018} identify such
filaments in thin channel maps with caustics caused by velocity
crowding. Following their assertion, the structures seen in
Figs. \ref{Fig_Gauss_0} and \ref{Fig_HFI_USM} need to be interpreted as
mere chance coincidences. In addition, the observed dependences between
the value of $I_{857}/N_{HI}$ and the strength of the USM signal cannot
be explained in this context. We show in Fig. \ref{Fig_Doppler} that the
USM and CNM structures are both cold, a finding that was also questioned
by \citet{Yuen2019}. Searching for an interpretation of the
$I_{857}/N_{HI}$ ratio in context with the warmer part of the multiphase
\hi\ distribution leads to considerable difficulties; we refer to the
most recent analysis of the large-scale distribution of interstellar
reddening derived from \hi\ emission by \citet{Lenz2017} and further
references discussed by these authors. The gas-to-dust ratio is well
defined only for $N_{\mathrm {HI}} \la 4 ~ 10^{20}$ cm$^{-2}$ and
$E(B-V) \la 0.08$ mag. Figure 9 of \citet{Lenz2017} shows that USM and
CNM structures in Figs. \ref{Fig_Gauss_0} and \ref{Fig_HFI_USM} belong
to a multiphase medium that is mostly out of this range. Variations in
the gas-to-dust ratio are a long-standing problem. \citet{Liszt2014}
attributed systematic trends in the gas-to-dust ratio to the onset of
\h2 formation at low temperatures. In view of these results our
argumentation became circular. Trying to explain the $I_{857}/N_{HI}$
ratio in context with the multiphase \hi\ distribution, we end up with
suggestions by \citet{Clark2019}, previously rejected by
\citet{Yuen2019}, that point to the coldest \hi\ components. VCA
basically considers a WNM dominated \hi\ fluid at Mach one, and the
properties of the embedded CNM are inconsistent with such an approach.

It remains to be explained why the velocity width dependent power
spectral indices derived by \citet{Stanimirovic2001} are in perfect
agreement with the VCA theory by \citet{Lazarian2000}. The derived power
spectra come from the observed \hi\ layer of the small Magellanic cloud,
extending roughly perpendicular to the line of sight on scales between
30 pc and 4 kpc. In our case, Fig. \ref{Fig_Plot_Power_CNM} shows excess
power for the CNM at $l \ga 400$, corresponding to scales $\la
0\fdg5$. This corresponds to the angular scale that was used by us to
filter USM structures, and corresponds to a linear scale of 1 pc at an
assumed distance of 100 pc. Filamentary \hi\ structures were previously
reported to have extensions well below a parsec (\citet{Clark2014} and
\citet{Kalberla2016}).  These are also scales where phase transitions
are expected (\citet{Audit2005}, \citet{Federrath2016}, and
\citet{Wareing2019}). Observations with a beam width of more than 30
times this scale cannot resolve phase fluctuations within the \hi\ layer
(imagine what kind of Milky Way ISM research you can do with a 3.3 m
dish in place of a 100 m dish). The power analysis for the SMC is
sensitive to fluctuations of the total \hi\ column density, to be
considered observationally as a single phase only. This is exactly the
condition assumed by \citet{Lazarian2000} and the findings of
\citet{Stanimirovic2001} provide a direct proof for VCA. In the case of
the Galactic \hi\ the single phase assumption is violated and VCA does
not apply. The statement by \citet{Yuen2019} that ``both two phase and
one phase show the same result'' is not supported by observations.
Fluctuations in spectral indices observed for \hi\ in the Milky Way are
predominantly affected by fluctuations in the phase composition. We
cannot exclude VCA predicted influences from velocity fluctuations, but
these must be marginal compared to changes in spectral index caused by
temperature dependences from the CNM. Accordingly velocity caustics must
be marginal, and observed filamentary \hi\ structures are caused
predominantly by dust-bearing density structures as proposed by
\citep{Clark2019}.

\section{Summary}
\label{Summary}

Intending to moderate the discussion between \citet{Clark2019} and
\citet{Yuen2019} on whether \hi\ filaments are dust-bearing density
structures or velocity caustics, we analyzed EBHIS data from the same
region around (GAL$ = 35\fdg5$, GAB$ = 54\deg$), as used by these
authors. We used unsharp masking to determine filamentary
\hi\ structures at high spatial frequencies. These structures are cold
and associated with FIR emission from dust at 857 GHz. We also used a
Gaussian decomposition to extract CNM structures. The USM emission is
found to be embedded within the CNM. While the FIR emission is
slightly more extended than the USM emission we find that the CNM
covers all FIR emission traced by the USM structures. We confirm the
results by \citet{Clark2019} that the $I_{857} / N_{HI}$ ratio is
elevated in regions with strong USM emission, implying that these
structures are density enhancements.

We test the VCA hypothesis that \hi\ filaments originate from velocity
caustics, caused by velocity crowding along the line of sight. We
calculated the power spectra for the FIR emission, and observed
\hi\ column densities and also for the CNM. The single-channel
\hi\ power spectra are steepest at velocities where the USM structures
and the CNM are coldest. At these velocities the 857 GHz power spectrum
($\gamma_{\mathrm{857}} \sim -2.58$) shows the best agreement with the
steepest CNM power spectrum ($\gamma_{\mathrm{CNM}} \sim -2.5$ at
$v_{\mathrm{LSR}} = -6 $ \kms). For single-channel power spectra we find
power index fluctuations up to $\delta \gamma \la 0.5$ in the case of
the multiphase \hi\ and $\delta \gamma \la 1$ in the case of the
CNM. These velocity dependences are large compared to the steepening
$\delta \gamma_{\mathrm{VCA}} = 0.08$ caused by a transition from thin
to thick velocity slices. The VCA predicted steepening is marginal in
comparison to thin slice index fluctuations that do not exist for an
isothermal medium.  We find that the spectral index for the multiphase
medium is steepest at those velocities where the CNM or USM emission is
coldest. Such a coincidence supports an explanation by phase
transitions. These lead to a local decrease in the thermal line widths,
but the column densities do not change significantly. Thus, thermal
instabilities produce enhanced power in the line centers of the observed
\hi\ at the expense of the power in the wings of the line. Thermal
instabilities occur predominantly at small scales, condensations from
WNM to CNM thus decrease the multiphase power predominantly at high
multipoles. In consequence, the multiphase power spectra steepen, a
process that was observed during simulations by \citet{Wareing2019}. CNM
power spectra are affected in the opposite sense. Phase transition at
small scales increase the power at high multipoles, leading to shallower
power spectra.

Our results, in agreement with \citep{Clark2019}, call for a significant
reassessment of many observational and theoretical studies of turbulence
in \hi, emphasizing in particular the response from phase
transitions. The VCA theory by \citet{Lazarian2000} is one of the most
important contributions to our understanding of the turbulent ISM and
has inspired hundreds of publications in this field. The reported
discrepancies between theory and observations hamper our understanding
of the relations between cold filamentary gas and dust and need urgently
be solved.

Comments and discussions in this paper are based on the version of
\citet{Yuen2019} as submitted on 5 Apr 2019 to arXiv, available at the
time of submission of our contribution. Currently there are 11 
citations registered by ADS; the conributions by \citet{Yuen2019} appear
to spread out quickly and we would like to argue against theoretical concepts
that are opposed by observations.

\begin{acknowledgements}
  P. K. thanks J{\"u}rgen Kerp for support and discussions.
  U. H. acknowledges the support by the Estonian Research Council grant
  IUT26-2, and by the European Regional Development Fund (TK133).  This
  research has made use of NASA's Astrophysics Data System. EBHIS is
  based on observations with the 100-m telescope of the MPIfR
  (Max-Planck-Institut f\"ur Radioastronomie) at Effelsberg. Some of the
  results in this paper have been derived using the HEALPix package. We
  also used the Karma package by R.E. Gooch. 
   \end{acknowledgements}

\begin{appendix} 
  \section{Citations concerning VCA} 
    \label{appA}

For the ease of the reader we collect here literal citations that are in
conflict with the presentation by \citet{Yuen2019}, and parameters as
far as their Table 1 is concerned.

\citet{Khalil2006} write in their abstract: 
``The slopes of the power spectra for an increasing number of velocity
channels were compared for 11 sections of the Local arm column density
mosaic. All slopes are identical within the uncertainties (-3.0) and we
do not detect for the Galactic plane the change in the power law index
predicted by Lazarian and Pogosyan.'' \citet{Khalil2006} give a detailed
discussion of these issues in their Sect. 6 and present the results of
their VCA analysis in their Tables 3 to 5, also in their Fig. 29.
All of these results are in conflict with citations by \citet{Yuen2019}
and their Table 1, column 1. 

\citet{Deshpande2000} write in their Sect. 5: ``In a recent paper,
\citet{Lazarian2000} draw attention to the modification of the H I
density spectrum due to the velocity field. They point out that the
velocity range needs to be thick in order to obtain the true density
power spectrum. They estimate a power-law index close to the Kolmogorov
value for the power spectrum of density using the \hi\ emission data of
the Galaxy \citep{Green1993} and of the Small Magellanic Cloud
\citep{Stanimirovic2001}.  Being aware of this complication, we have
obtained the power spectra before (case A) and after (case B) averaging
in velocity (§4). Within the errors of estimation, the two slopes agree
(Fig. 6), indicating that the effect of averaging over velocity is not a
substantial effect.''

\citet{Choudhuri2018} write in their abstract: ``We also measure the
power spectra after smoothing the 21 cm emission to velocity resolution
ranging from 1.03 to 13.39 \kms, but the power spectrum remains
unchanged within the error bar.'' In their Sect. 3 they present in
Fig. 7 their data after averaging different velocity channels and
conclude: ``Clearly, we do not observe any significant difference in the
power spectrum after velocity channel averaging. This signifies that the
fluctuations are mainly dominated by the density fluctuations.''

\section{Related biliographic entries disappear in the arXiv version of a nature paper} 
\label{appB}

After receiving the first referee report on the present paper we
resubmitted on 15 February 2020 a revision of our manuscript that was
else unpublished at this date.  A few days later, on 23 February 2020,
the publication by \citet{Hu2019b} appeared as the arXiv version of the
original nature astronomy publication \citep[][10 June
  2019]{Hu2019a}. Whether both events are related to each other is not
clear; the coincidence is strange but we may be missing something in
this story. Surprisingly, concerning the controversy between
\citet{Clark2019} and \citet{Yuen2019}, discussed by us in detail, we
note some discrepancies between both versions of the Hu et
al. publications.

\citet{Hu2019b} state in their Methods section: ``The issue of whether the
small scale structures in neutral hydrogen velocity channel maps are
dominated by density or velocity structures has been debated recently,
preprints by Clark et al. (2019)$^{\color{blue}{53}}$ and the response
in Yuen et al. (2019)$^{\color{blue}{54}}$. However, irrespectively of
the outcome of these debates, our conclusion that the velocity channel
gradients trace magnetic fields well is not affected, especially in the
regime of molecular clouds.''

While the text is identical in both versions we find in the arXiv
version that the nature-style references $^{\color{blue}{53}}$ and
$^{\color{blue}{54}}$ are dummies, the original bibliographic entries 53
and 54 are missing and the entries by number refer in \citet{Hu2019b} to
completely different unrelated publications. The same happened to
references 51 and 52 that refer to the imprint of turbulence on PPV
data. Inspecting the arXiv source file shows that in all four cases the
modifications were generated through replacing
``\char`\\cite\char`\{...\char`\}'' by
``\char`\$\char`\^\char`\{\char`\\color\char`\{blue\char`\}\char`\{...\char`\}\char`\$''.
These constructs allow to generate pretended nature-style links to
references without the need of existing valid destinations when
compiling the source file. In the \citet{Hu2019a} version, published in
nature astronomy, all bibliographic entries have valid references.

ArXiv does not care about consistency of the submission. Nature
astronomy asks authors to post the submitted version but does not take
responsibility for versions on arXiv. 

\end{appendix}

\end{document}